\documentclass[referee]{raa} 

\usepackage{graphicx,times}
\usepackage{natbib}
\usepackage{multirow}
\usepackage{amsmath}
\usepackage{booktabs}
\usepackage{rotating}
\usepackage[pagebackref=true]{hyperref}

\newcommand{\degc}{^{\circ}}

\usepackage{color}

\begin{document}
   \title{Detection of Quasi-periodic Oscillations in SGR 150228213}

 \volnopage{ {\bf 20XX} Vol.\ {\bf X} No. {\bf XX}, 000--000}
   \setcounter{page}{1}

   \author{Run-Chao Chen\inst{1}, Can-Min Deng\inst{1}, Xiang-Gao Wang\inst{1}, Zi-Min Zhou\inst{1}, {Xing Yang}\inst{1}, Da-Bin Lin\inst{1}, Qi Wang\inst{1}, En-Wei Liang\inst{1}
   }

   \institute{Guangxi Key Laboratory for Relativistic Astrophysics, School of Physical Science and Technology, Guangxi University, Nanning 530004, China; {\it dengcm@gxu.edu.cn}; {\it wangxg@gxu.edu.cn}
\vs \no
   {\small Received 20XX Month Day; accepted 20XX Month Day}
}

\abstract{
The detection of quasi-periodic oscillations (QPOs) in magnetar giant flares (GFs) has brought a new perspective to study the mechanism of magnetar bursts.
Due to the scarcity of GFs, searching QPOs from magnetar short bursts is reasonable.
Here we report the detection of a high frequency QPO at approximately 110 Hz and a wide QPO at approximately 60 Hz in a short magnetar burst SGR 150228213, with a confidence level of 3.35$\sigma$. 
This burst was initially attributed to 4U 0142+61 by $Fermi$/GBM on location, but we haven't detected such QPOs in other bursts from this magnetar.
We also found that there was a repeating fast radio burst associated with SGR 150228213 on location.
Finally, we discuss the possible origins of SGR 150228213.
}

   \authorrunning{R.-C. Chen et al. }     
   \maketitle

%________________________________________________ sections below
%  
\section{Introduction} \label{sec:intro}
Magnetars are a class of young neutron stars that have the strongest magnetic fields in the universe so far.
They have typical magnetic fields $B \sim 10^{14}$ G, spin period $P \sim 2$$-$$12$ s and spin down rate $\dot{P} \sim 10^{-13}$$-$$10^{-11}$ s s$^{-1}$ \citep{2015RPPh...78k6901T}.
These isolated neutron stars emitted a wide array of electromagnetic radiation in radio, optical, X-ray and gamma-ray band by the decay of their enormous internal magnetic fields, which also brings the name 'magnetar' \citep{2017ARA&A..55..261K,1992ApJ...392L...9D}.
Magnetars can be divided into Soft Gamma Repeaters (SGRs) and Anomalous X-ray Pulsars (AXPs) judging from burst activities and other aspects.

Bursts from magnetars can be divided into three categories: short bursts is the most common type which has typical duration $\sim$ 0.1 s and peak luminosities $\sim$ $10^{39}$$-$$10^{41}$ erg s$^{-1}$; intermediate flares are rare events that usually last 1$-$40 s with peak luminosities $\sim$ $10^{41}$$-$$10^{43}$ erg s$^{-1}$; GFs are the most violent and unique activities in magnetars, which have an extremely bright hard peak last 0.1$-$0.2 s with a luminosity of $\rm 10^ {44} $$-$$10^ {47} $ erg s$^ {-1} $, usually followed by a long pulsating tail lasting a few hundred seconds modulated by the magnetar spin period \citep{2015RPPh...78k6901T}, only four events were confirmed (GRB 790305 from SGR 0526-66 \citep{1979Natur.282..587M,1980ApJ...237L...1C}, GRB 980827 from SGR 1900+14 \citep{1999Natur.397...41H,1999ApJ...515L...9F,1999AstL...25..635M}, GRB 041227 from SGR 1806-20 \citep{2005Natur.434.1098H,2005Natur.434.1104G,2005Natur.434.1107P,2005Natur.434.1112C} and GRB 200415A \citep{2020ApJ...899..106Y,2020ApJ...903L..32Z,2021Natur.589..211S,2021Natur.589..207R}).

The association event of SGR 1935+2154$-$FRB 200428 on 28th April 2020 \citep{2020Natur.587...54C,2020Natur.587...59B,2020Natur.587...63L,2021NatAs...5..378L,2020ApJ...898L..29M,2021NatAs...5..372R} had established that at least some fast radio bursts (FRBs) are produced during magnetar bursts \citep{2014MNRAS.442L...9L,2016ApJ...826..226K,2018ApJ...868...31Y,2021Univ....7...56L,2021MNRAS.500.2704Y}, but the mechanism behind this phenomena is unclear.
Starquakes have been invoked to explain the occurring of hard X-ray bursts and FRBs from magnetars \citep{1995MNRAS.275..255T,2018ApJ...852..140W}.
This kind of crustal oscillations would leave imprints in the form of QPOs in the temporal profiles of magnetar bursts \citep{2014ApJ...793..129H,2019ApJ...871...95M}.

The QPOs have been found during the pulsating tails and the main peak of magnetar GFs \citep{1983A&A...126..400B,2005ApJ...632L.111S,2005ApJ...628L..53I,2006ApJ...653..593S,2021Natur.600..621C}, and also have been found in some short bursts from SGRs \citep{2014ApJ...795..114H,2014ApJ...787..128H,2022ApJ...931...56L}.
These investigations have opened the possibility of magnetar studying using asteroseismology \citep{2013ApJ...768...87H}.
At present, due to the scarcity of GFs, searching QPOs from short bursts is reasonable, although the duration of short bursts would limit the minimum frequency for QPOs searching \citep{2013ApJ...768...87H}.
In this paper we conduct a comprehensive analysis of SGR 150228213 and report the (quasi-)periodic signal detection in this burst.
The structure of this paper is as follows.
In Section \ref{sec:pds} we describe the Bayesian framework for searching (quasi-)periodic signals in the observed periodogram of magnetar bursts and estimating the significance.
Section \ref{sec:1502} is the periodogram analysis of SGR 150228213, we stated how to select samples and choose the appropriate time interval to conduct such analysis.
We also discussed the results of (quasi-)periodic research at this section.
In Section \ref{sec:discuss} we discussed origins of SGR 150228213 and Section \ref{sec:sum} is a summary to this work.

\section {Methods for periodogram analysis} \label{sec:pds}
\subsection {Generate the periodogram}\label{sec:pid}
The observed periodogram analyzed in this work is based on Fast Fourier Transform (FFT) of the light curve data from the selected time interval.
Powers in observed periodogram are corresponding to the squared Fourier transform of the data, and we make use of the $stingray$ python package \citep{2019ApJ...881...39H} to perform this conversion to get the Leahy-normalized periodograms.

Periodogram generated from pure noise process can be seen as the conversion of a stochastic time series.
It is well known that the periodogram of any stochastic time series of length $N$, denoted $I_j = I (f_j)$ at Fourier frequency $f_j = j/N{\Delta}T$ (with j = 1, ..., N/2), is exponentially distributed about the true spectral density $S_j = S(f_j)$
\begin{equation}
	p(I_j|S_j) = \frac{1}{S_j}exp(-I_j/S_j)
\end{equation}
\citep{1975ApJS...29..285G,1983ApJ...266..160L,1995A&A...300..707T}.
Thus we sampled the exponential distribution corresponding to the model power to generate (see \cite{2010MNRAS.402..307V}) the simulated periodograms in this work.

\subsection {Model the periodogram}
There are two alternative approaches to model the periodogram, one is relying on the light curve models to the original light curve to generate the periodogram, and another is directly using the models of the observed periodogram.
Modeling the original light curve is based on an accurate understanding of the burst mechanism, otherwise artificial model selection would bring an immeasurable impact on potential QPO detection.
Owing to the unknown emission mechanism of magnetar bursts, we chose to model the observed periodogram generated from the original light curve to search for (quasi-)periodic signals in magnetar bursts.

While modeling the observed periodogram, we made a simple but conservative assumption that all broadband powers in periodogram is supplied by a noise process without QPO, which is the combination of red-noise at low frequencies and white-noise at high frequencies \citep{2013ApJ...768...87H}.
Based on this assumption, searching for (quasi-)periodic signals through the periodogram research can be followed by the Bayesian approach developed by \cite{2010MNRAS.402..307V}, such method provides a statistically rigorous framework to test whether additional model components (such as Lorentzian QPOs) are required by the data.
And as was stated in \cite{2021Natur.600..621C}, such assumption will cause weak signals at low frequencies to be buried in the higher variance of the broadband noise but would yield a very low false positive detection rate in return.

A theoretical pure red-noise profile follows a broken power-law model, but in many cases the break frequency is relatively small, and the red-noise profile would be fitted better by the power-law model \citep{1992ApJ...393..266B,2002MNRAS.337.1426L}.
Therefore, we need to select the preferred noise model of the observed periodogram from these two nested models below.
We defined the PL model as a red-noise power-law function plus a white-noise (Poisson noise) constant as
\begin{equation}
P(\nu) = A \nu ^ {-\alpha} + C,
\end{equation}
where $\nu$ is the frequency, $P(\nu)$ is the power, A is the amplitude, $\alpha$ is the power-law index and C is the constant representing white-noise level.
And the BPL model is the combination of a broken power-law and a white-noise constant, which is described as
\begin{equation}
P(\nu) = N \left[ 1 + \left( \frac{\nu}{\nu_b} \right) ^{\beta} \right] ^{-1} + C,
\end{equation}
where N is the normalization value, $\nu_b$ is the break frequency and $\beta$ is the power-law index after $\nu_b$.

As for the model fitting, we obtained the optimum model parameter set from the maximum a posteriori (MAP) estimates, which could be computed by minimizing the maximum likelihood estimation (MLE) function \citep{2010MNRAS.402..307V,2013ApJ...768...87H}
\begin{equation}
D(\textbf{\emph{I}}, \bm{\theta}, H) = -2\log p(\textbf{\emph{I}}| \bm{\theta}, H) = 2 \sum_{j=1}^{N/2} \left\{ \frac{I_j}{S_j} + \log S_j \right\},
\end{equation}
where $p(\textbf{\emph{I}}| \bm{\theta}, H)\,=\,\prod_{j}^{N/2}p(I_j|S_j)$ is the joint likelihood function, $I_j$ is the individuals power in observed periodogram and $S_j$ is the power in noise model for a parameter set $\bm{\theta}$.

To select the preferred noise model, we make use of the likelihood ratio test (LRT).
The $null$ $hypothesis$ is that the periodogram can be described by a simple model, PL ($H_0$), then we estimated whether the $H_0$ model could be replaced by a more complex model, BPL (the $alternative$ $hypothesis$, $H_1$) through the LRT statistic
\begin{equation}
\begin{aligned}
T_{\rm LRT} &= -2\log\frac{p(\textbf{\emph{I}}| \hat{\bm{\theta}}^0_{\rm MLE}, H_0)}{p(\textbf{\emph{I}}| \hat{\bm{\theta}}^1_{\rm MLE}, H_1)} \\
&= D_{\min}(H_0) - D_{\min}(H_1).
\end{aligned}	
\end{equation}
We can generate $n$ sets of simulated periodograms by sampling the posterior distribution of $H_0$ model parameters, we then compute the corresponding $T_{\rm LRT}$ by fitting each fake periodogram with both $H_0$ and $H_1$ model.
The preferred noise model can be judged from the tail area probability ($p$$-$value) of the observed $T_{\rm LRT}$ in the distribution of the simulated $T_{\rm LRT}$.
It is necessary to emphasize that this test cannot be seen as direct evidence in favor of the $H_1$ model (usually the more complex one), but only a strictly evidence against the $H_0$ model \citep{2014ApJ...787..128H}.

\subsection {Search for (quasi-)periodic signals}\label{sec:period}
After the selection of noise model, we use the preferred noise model to search for periodic signals or QPO candidates.
We computed residuals of the observed power to noise model power in the logarithmic periodogram from the selected noise model with the optimum parameter set.
Such residual is equivalent to $I_j/S_j$, for which we can use the $T_R$ statistic to estimate the chance probability of the candidates.
$T_R$ is the maximum ratio of observed to model power described as
\begin{equation}
T_R = \max(\hat{R_j}),
\end{equation}
where
\begin{equation}
\hat{R_j} = 2I_j/S_j.
\end{equation}
In this step, we generated $n$ sets of simulated periodograms by sampling the posterior distribution of the selected noise model parameters, from each periodogram we could obtain the new $T_R$.
These statistics would be distributed as $\chi^2$ and we can get the $p$-value of $T_R$ by computing the tail area probability \citep{2010MNRAS.402..307V,2013ApJ...768...87H}.

As for searching for QPOs in observed periodogram, it is similar to the selection of noise models.
In this step, the $null$ $hypothesis$, $H_0$, becomes that the periodogram can be well described by the selected noise model, and the $alternative$ $hypothesis$, $H_1$, model is the superposition of $H_0$ noise model and one or several Lorentz lines account for QPOs \citep{2021Natur.600..621C}.
The Lorentz line is described as \citep{1996ASPC..101...17A}
\begin{equation}
P(\nu) = K(\sigma/2\pi)/[(\nu-\nu_p)^2+(\sigma/2)^2],
\end{equation}
where K is the normalization factor, $\sigma$ is the FWHM (full width at half maximum) of the line and $\nu_p$ is the centroid frequency of QPO.
We can take the $p$$-$value of $H_1$ model obtained by LRT statistics as the significance of such QPO based on the establishment of $H_0$.

\section {Periodogram analysis for SGR 150228213}\label{sec:1502}
\subsection {Sample selection}
$Fermi$/GBM is an all-sky monitor for any burst event and covering the energy range from 8 keV$-$40 MeV \citep{2009ApJ...702..791M}, which is suitable for the detection of short bursts from magnetars.
After years of accumulation, we have collected 524 bursts information which was classified as SGR by machine from the official website of $Fermi$\footnote{https://fermi.gsfc.nasa.gov}, 177 of them is certified from the known sources and other 347 bursts are certified from unknown sources.

Studies for magnetar bursts based on the observation data from $Fermi$/GBM usually target specific magnetars for batch analysis and especially for those active SGRs e.g., SGR 1935+2154 \citep{2020ApJ...893..156L}, SGR 1550-5418 \citep{2014ApJ...787..128H}.
In this work, we focused on those bursts which were certified from unknown sources and preferred those associated with known magnetars or FRBs (sources), for which are more likely to be originated from magnetars.
Therefore, we compared the location information for the 347 bursts from unknown sources with 30 magnetars form the McGill Magnetar Catalog\footnote{https://www.physics.mcgill.ca/~pulsar/magnetar/main.html} \citep{2014ApJS..212....6O}, and 626 FRBs from the CHIME/FRB catalog\footnote{https://www.chime-frb.ca} \citep{2021arXiv210604352T}, FRBCAT\footnote{https://www.frbcat.org} \citep{2016PASA...33...45P} and TNS\footnote{https://www.wis-tns.org}.
After the comparison, except for the SGR 1935+2154 associated samples, we found only one burst, SGR 150228213, is related to a known magnetar, AXP 4U 0142+61, and a repeating FRB source, FRB 180916, on location.

However, 4U 0142+61 is not associated with FRB 180916, but the periodogram analysis for SGR 150228213 has revealed a possible periodic or quasi-periodic signal.
The later content of this chapter describes our periodogram analysis for SGR 150228213 and the significance estimation of related results, and the discussion of two different origins will be carried out in Chapter \ref{sec:discuss}.

\subsection {Temporal analysis}
Considering that short bursts from magnetars usually have short duration and soft energy spectrum, we combined the Time-Tagged Event (TTE) data files from all triggered NaI detectors (n4, n8) and rebinned the data in 2 ms time resolution to analyze the light curve in energy range of 8$-$100 keV.

\begin{figure}
	\centering\includegraphics[width=\linewidth]{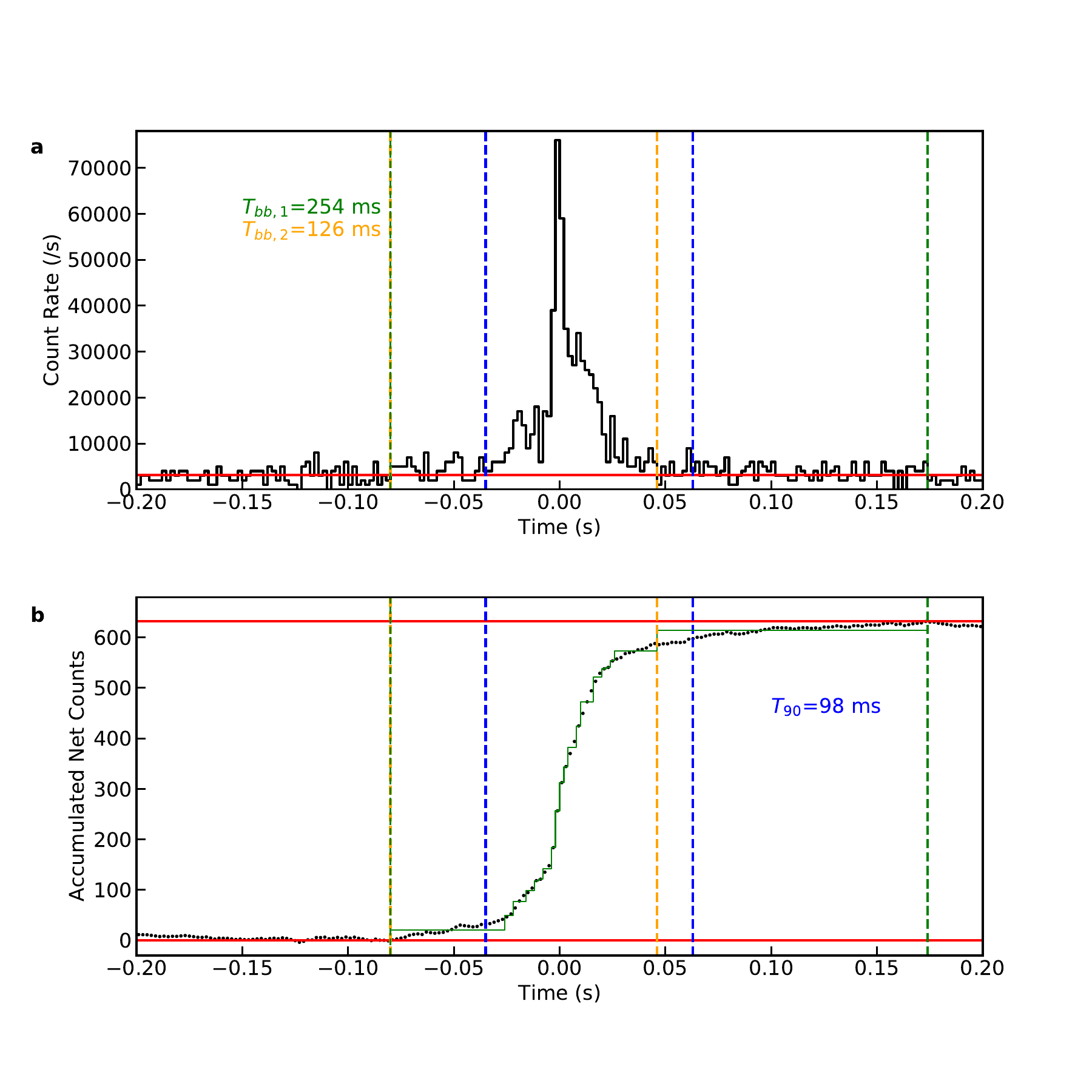} 
	\caption{Light curve of SGR 150228213. \textbf{a}, the black solid line represents the light curve obtained by the combination of events data from detector n4 and n8 in 8$-$100 keV. The red solid line shows the background level. \textbf{b}, the black points represent the accumulated counts variation, the red solid lines show the 0$\%$ and 100$\%$ level of the total accumulated counts. The two regions marked by the blue and yellow vertical dashed lines are the $T_{90}$ and Bayesian Block Time ($T_{bb, 2}$) intervals of the light curve, the time interval for $T_{bb, 1}$ within green vertical dashed lines is the total burst duration we select to conduct the periodogram research.}
	\label{lc}
\end{figure}

We use $T_{90}$ to describe the main part of this burst, which is the time interval within the accumulated counts of the burst increases from 5$\%$$-$95$\%$ of the total counts \citep{1993ApJ...413L.101K}.
Since the estimation of $T_{90}$ will be affected by background level, and SGR 150228213 was triggered during the active phase of 4U 0142+61, which had made the background fluctuated greatly, we selected a relatively long time interval near the burst to estimate the average background level to neutralize the effects of some potentially weak bursts, which is the time intervals of -25$-$-1 s and 1$-$25 s to the trigger time $T_0$.
The light curve within -0.2$-$0.2 s is shown in Figure \ref{lc}, in which we also drew the net counts accumulation diagram corresponding to the light curve, the $T_{90}$ we computed is $\sim$ 98 ms.

In addition, to depict the local characteristics of the burst and select the suitable interval to conduct the periodogram research, we adopted the Bayesian Blocks algorithm described in \cite{2013ApJ...764..167S} to analyze the light curve data within -25$-$25 s to $T_0$ in the same time resolution.
The accumulated net counts of each 'blocks' were also drawn in Figure \ref{lc} in the form of a ladder graph.
For the light curve in Figure \ref{lc} which includes the total duration to calculate $T_{90}$ of the burst, there is a long 'block' between the 95$\%$$-$100$\%$ interval of the total accumulated net counts.
According to the description in \cite{2021ApJ...906L..12Y}, the Bayesian block duration time $T_{bb}$ for bursts from magnetar SGR 1935+2156 has a power-law trend with $T_{90}$ in $T_{90} \propto T^{0.91\pm0.05}_{bb}$. 
Following this correlation, $T_{bb, 2}$ $\sim$ 126 ms in Figure \ref{lc} is the suitable Bayesian block duration of SGR 150228213.
However, since the interval of $T_{bb, 2}$ does not contain the main part of the burst ($T_{90}$), we treat the interval of $T_{bb, 1}$ (-80$-$174 ms to $T_0$) as the total duration of SGR 150228213 for periodogram analysis, which contains the main part of the burst and the part which could not be distinguished from the burst or background.

\subsection {Periodogram analysis}\label{sec:qpo}
According to temporal analysis of SGR 150228213, we select two different time segments to compute the observed periodograms: the interval of -80$-$174 ms to $T_0$ denotes the total duration of the burst based on Bayesian blocks, and the interval of -44.8$-$72.8 ms to $T_0$ is the interval of $T_{90}$ with 20$\%$ exceed part refers to \cite{2013ApJ...768...87H} (which also denotes the total burst based on $T_{90}$).
We combined the event data from all detectors in 8-100 keV and rebinned the light curve data in 0.2 ms time resolution (corresponding to a Nyquist frequency of 2500 Hz).

Referring to \cite{2014ApJ...787..128H}, the specific process for noise model selection and LRT statistic is as follows.
\begin{enumerate}
\item We make use of the $emcee$ python package \citep{2013PASP..125..306F} to perform a suit of Markov chain Monte Carlo simulations (MCMCs) and sampled the posterior predictive distribution of the $H_0$ model (PL) with 50 MCMC ensemble walkers and 1000 samples for each walker (containing 20$\%$ samples in burn-in phase for each walker).
\item We simulated 1000 sets of periodogram from the MCMC sample of PL model and fit each periodogram with PL and BPL model to compute the distribution of $T_{\rm LRT}$ for those fake periodograms.
\item If the $p$$-$value of rejecting the PL model ($H_0$) from the observed periodogram falls below 0.05, we selected the BPL model as the preferred noise model. Otherwise, we preserve the PL model as the preferred noise model.
\end{enumerate}
After the selection of noise model, we found that the preferred noise model of both segments are PL.
We then use the PL model with optimum parameter set to calculate a boundary frequency of the red-noise dominated part to white-noise dominated part through ${\nu}=(A/C)^{(1/{\alpha})}$.
Divided by this boundary, we can compute $T_R$ in each part on the observed periodogram and obtain the corresponding (quasi-)periodic candidates.
Using the MCMC sample of PL model, we simulated 1000 sets of periodograms to compute the distribution of $T_R$ in each part on the fake periodograms and then estimate the corresponding $p$$-$value of each candidate.

Results for noise model selection and periodic research in different time segments are presented in Table \ref{pds_results1}.
It can be seen from the results that there might be a possible periodic signal or QPO candidate $\sim$ 110 Hz at each observed periodogram, which is located within the red-noise dominated part.
The signal with minimum $p$($T_R$) appears at the time interval of -80$-$174 ms.

\begin{table*}
	\caption{The preferred noise model and potential periodicities in SGR 150228213}
	\label{pds_results1}
	\begin{center}
		\begin{tabular}{cccccccc}
			\hline\hline
			Time Interval & \multicolumn{4}{c}{Noise Model Selection} & \multicolumn{3}{c}{Search for periodicties} \\
			\cmidrule(r){2-5}\cmidrule(r){6-8}
			(ms) & Model & $T^{\rm BPL}_{\rm LRT}$ & $p$(LRT) & Boundary (Hz) & Frequency (Hz) & $T_R$ & $p(T_R)$ \\
			\hline
			\multirow{2}{*}{-80$-$174} & \multirow{2}{*}{PL} & \multirow{2}{*}{-5.22} & \multirow{2}{*}{0.881} & \multirow{2}{*}{118.22} & 114.17 & 13.13 & 0.021 \\ 
			&&&& & 1590.55 & 11.55 & 0.853 \\
			\hline
			\multirow{2}{*}{-44.8$-$72.8} & \multirow{2}{*}{PL} & \multirow{2}{*}{-5.55} & \multirow{2}{*}{0.947} & \multirow{2}{*}{159.88} & 110.54 & 9.98 & 0.075 \\ 
			&&&& & 1590.14 & 11.81 & 0.522 \\
			\hline		
		\end{tabular}	
	\end{center}
\end{table*}

Considering these candidates could be a narrow QPO signal at $\sim$ 110 Hz, we add one Lorentz line to PL model as new $H_1$ model to fit the observed periodograms in each time segment.
Frequency of each QPO candidate is set as the initial value of the centroid frequency of the Lorentz line, and the width of this QPO was limited within a very narrow range (less than three times of the minimum frequency in each periodogram).
As can be seen from Table \ref{pds_results2}, the centroid frequency of this narrow QPO we suspect in all segments is still at about 110 Hz.
We then drew 5000 sets of simulated periodograms from the MCMC sample of PL model (new $H_0$ for QPO research) to compute the distribution of LRT statistics from PL and PL+QPO models, such QPO with the lowest $p$$-$value (the tail area fraction of $T^{obs}_{\rm LRT}$) $\sim$ 0.0008 exists in -80$-$174 ms interval, which is also consistent with the result above.

\begin{figure}
	\centering
	\includegraphics[angle=0,width=0.48\linewidth]{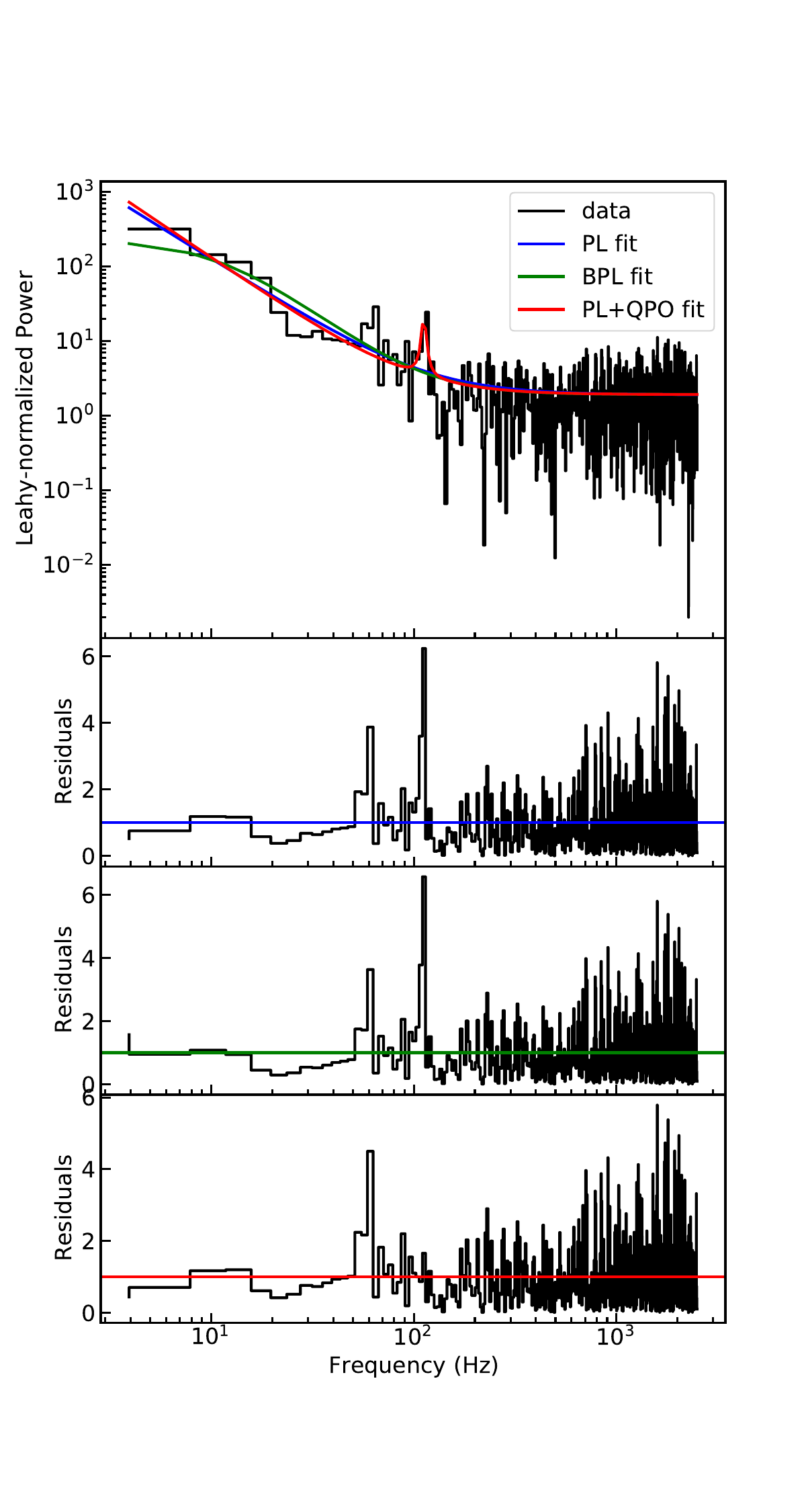}
	\includegraphics[angle=0,width=0.48\linewidth]{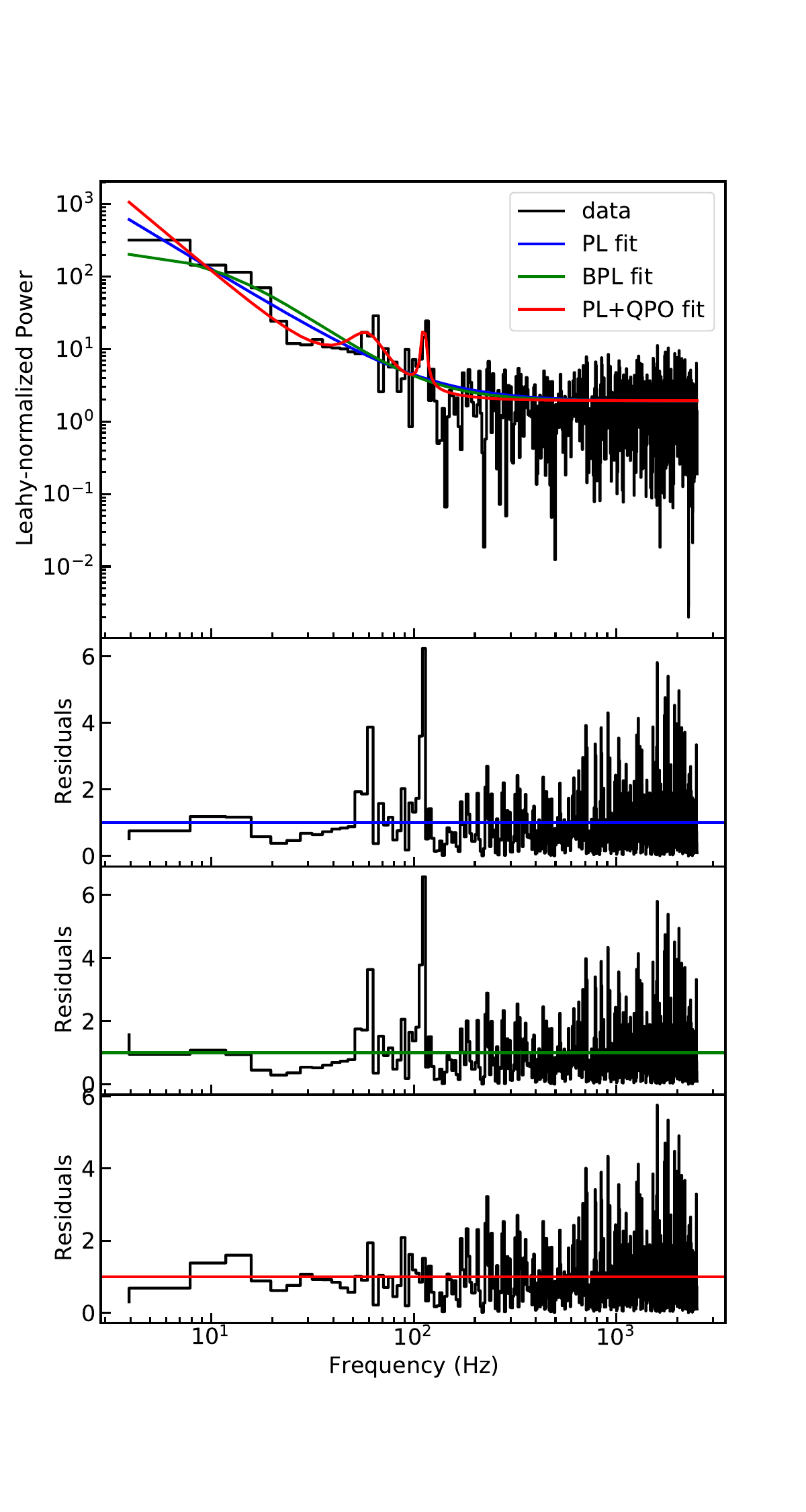}
	\caption{The observed periodogram from time interval of -80$-$174 ms to $T_0$. The left panel presents the diagram for QPO model based on the assumption that the periodic signal is a potential QPO signal. The right panel presents the diagram for QPO model of a wide QPO at 57.54 Hz and a narrow QPO at 112.20 Hz.}
	\label{PDS1}
\end{figure}

\begin{figure}
	\centering
	\includegraphics[angle=0,width=0.48\linewidth]{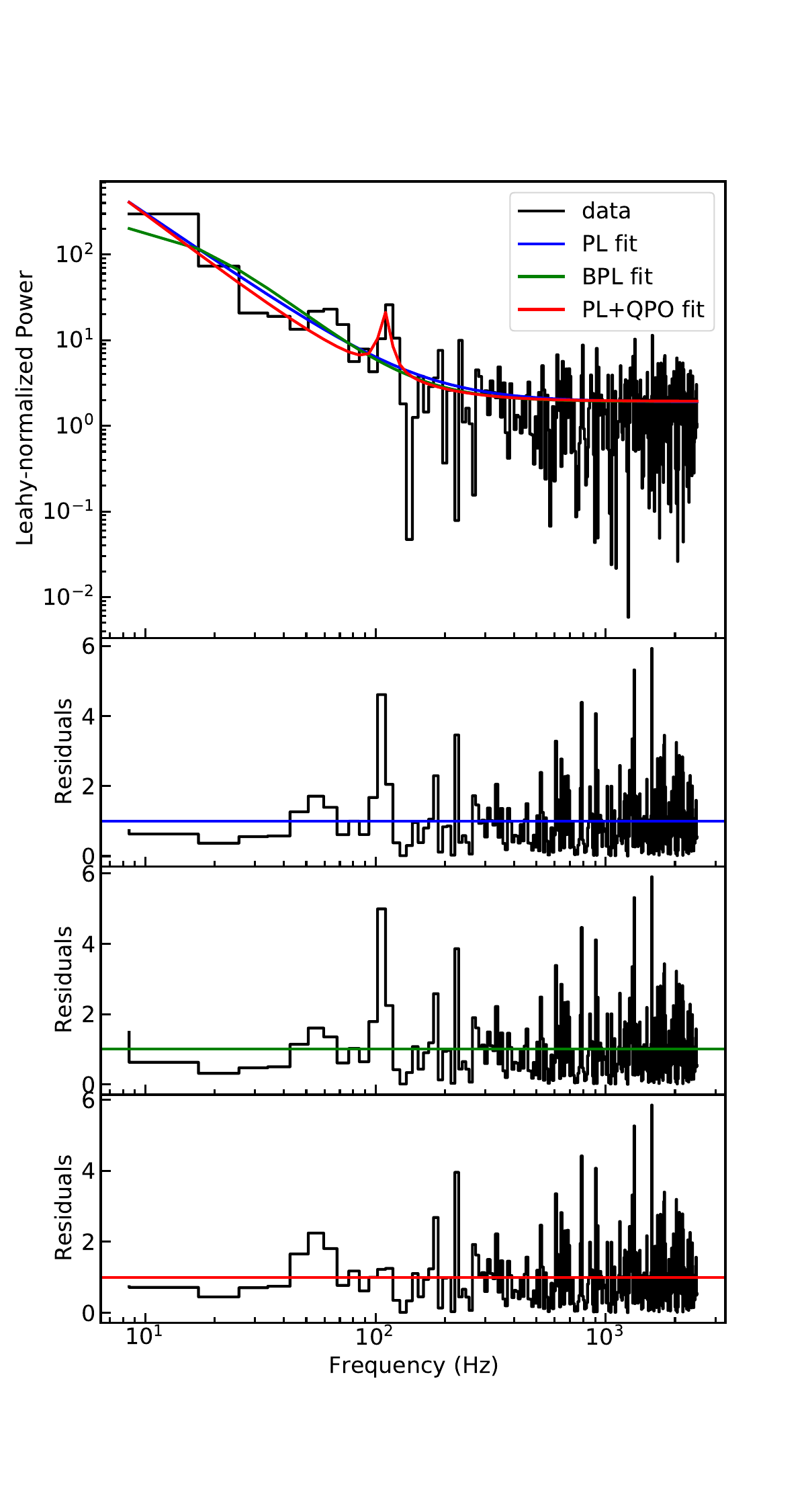}
	\includegraphics[angle=0,width=0.48\linewidth]{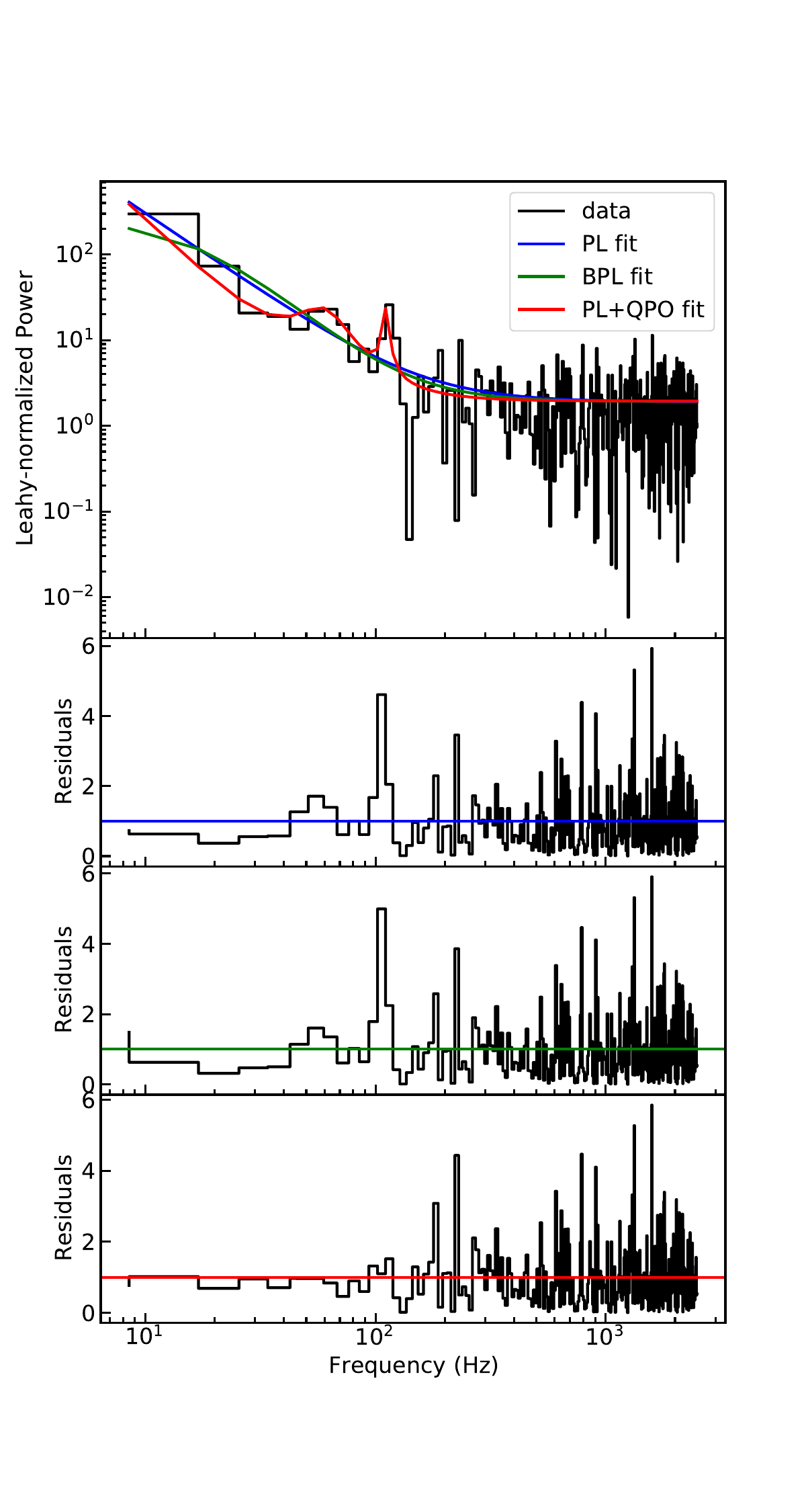}
	\caption{The observed periodogram from time interval of -44.8$-$72.8 ms to $T_0$. Both panels present a QPO at about 110 Hz, and the right panel presents the possible wide QPO at about 60 Hz.}
	\label{PDS2}
\end{figure}

Figure \ref{PDS1} is the periodogram of the observed data in -80$-$174 ms to $T_0$ and the corresponding models with the optimum parameter sets in each step of the periodogram analysis, we noticed that there is still exist a potential wide QPO signal at about 60 Hz.
However, such signal is not significant enough at the periodogram in -44.8$-$72.8 ms to $T_0$ (Figure \ref{PDS2}).
In order to find this potential wide QPO, we continued to use the QPO model with two Lorentz lines as new $H_1$ model to fit the observed periodograms.
In this case, we no longer restrict the width parameter for the wide QPO component but still set its initial centroid frequency at 110 Hz.
We still use 5000 sets of simulated periodograms generated from the MCMC sample of PL model to compute the distribution of LRT statistics and estimate the corresponding $p$$-$value of the new $H_1$ model with a wide QPO and a narrow QPO.
The fitting results corresponding to each time segment are presented in Table \ref{pds_results2}.
We can see that the frequency of the narrow QPO is still $\sim$ 110 Hz in time interval of -80$-$174 and -44.8$-$72.8 ms, and the wide QPO component locate at approximately 60 Hz.
The results with lowest $p$$-$value $\sim$ 0.0004 still exists in -80$-$174 ms interval.

\begin{table*}
	\caption{Parameter posteriors and chance probabilities for potential QPOs in SGR 150228213}
	\label{pds_results2}
	\begin{center}
		\begin{tabular}{ccccccccc}
			\hline\hline
			Time Interval & \multicolumn{2}{c}{Noise Model} & \multicolumn{6}{c}{Search for QPOs} \\
			\cmidrule(r){2-3}\cmidrule(r){4-9}
			(ms) & ($H_0$) & $D_{\min}$($H_0$) & Frequency (Hz) & FWHM (Hz) & Norm & $D_{\min}$($H_1$) & $T^{obs}_{\rm LRT}$ & $p$(LRT)  \\
			\hline
			\multirow{3}{*}{-80$-$174} & \multirow{3}{*}{PL} & \multirow{3}{*}{1700.56} & $112.20^{+1.02}_{-1.02}$ & $5.64^{+3.95}_{-3.50}$ & $162.18^{+2.29}_{-2.14}$ & 1694.44 & 6.12 & 0.0008 \\ 
			\cmidrule(r){4-9}
			&&& $57.54^{+1.12}_{-1.15}$ & $26.84^{+31.36}_{-17.41}$ & $594.54^{+1.58}_{-1.45}$ & \multirow{2}{*}{1691.94} & \multirow{2}{*}{8.63} & \multirow{2}{*}{0.0004}	\\	
			&&& $112.20^{+1.02}_{-1.02}$ & $4.80^{+4.49}_{-3.07}$ & $173.78^{+2.13}_{-2.40}$ &&& \\
			\hline
			\multirow{3}{*}{-44.8$-$72.8} & \multirow{3}{*}{PL} & \multirow{3}{*}{791.82} & $109.64^{+1.05}_{-1.05}$ & $10.98^{+9.43}_{-6.98}$ & $295.12^{+2.40}_{-2.95}$ & 788.42 & 3.29 & 0.0058 \\ 
			\cmidrule(r){4-9}
			&&& $61.66^{+1.29}_{-1.23}$ & $33.40^{+54.72}_{-22.13}$ & $1000.00^{+1.86}_{-2.14}$ & \multirow{2}{*}{790.05} & \multirow{2}{*}{1.77} & \multirow{2}{*}{0.0032}	\\	
			&&& $109.65^{+1.05}_{-1.05}$ & $7.21^{+3.22}_{-4.36}$ & $218.78^{+2.57}_{-3.02}$ &&& \\
			\hline		
		\end{tabular}	
	\end{center}
\end{table*}

\subsection {Duration of the QPOs}\label{sec:wavelet}
\begin{figure}
	\centering
	\includegraphics[width=0.48\linewidth]{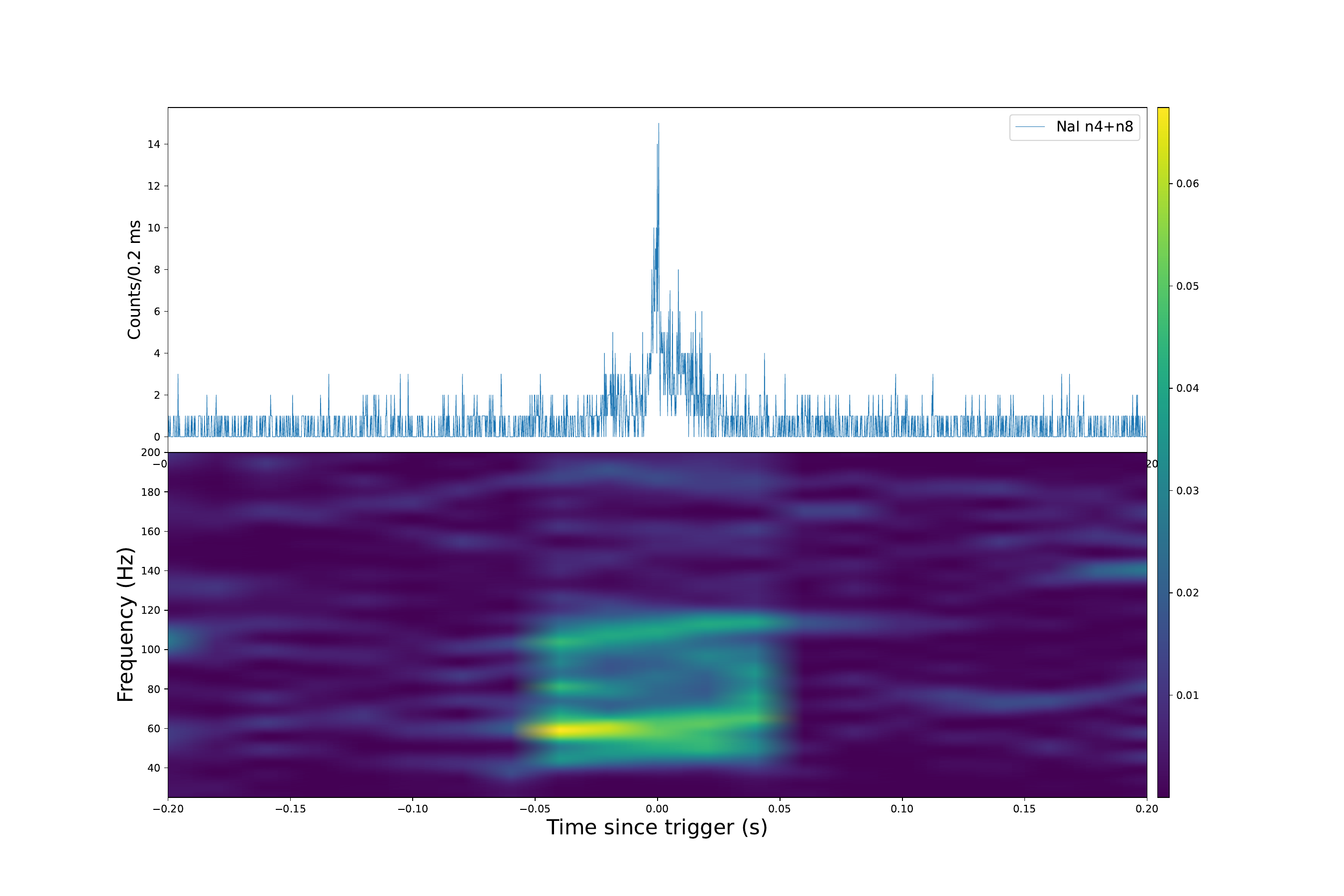} 
	\includegraphics[width=0.48\linewidth]{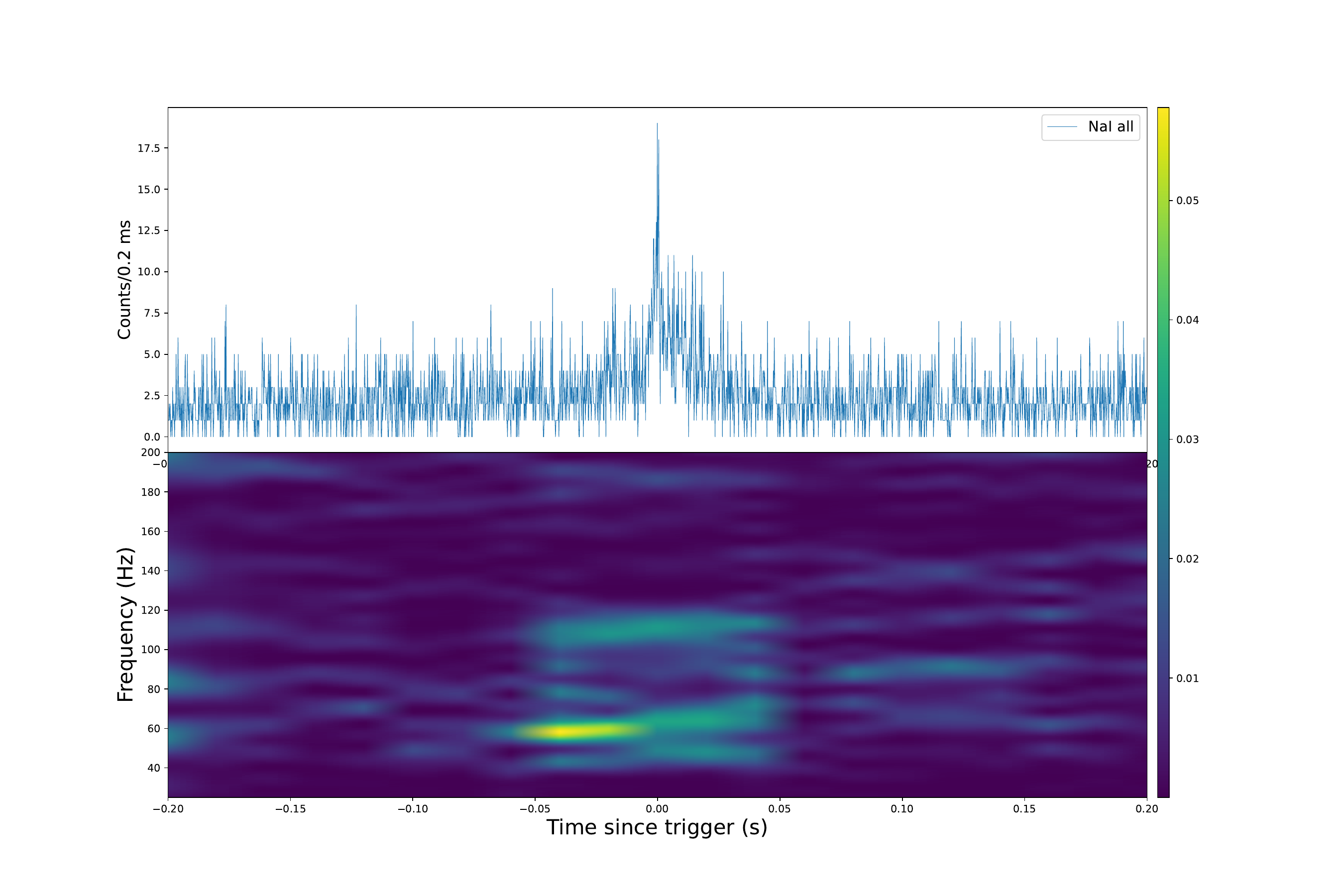}
	\includegraphics[width=0.48\linewidth]{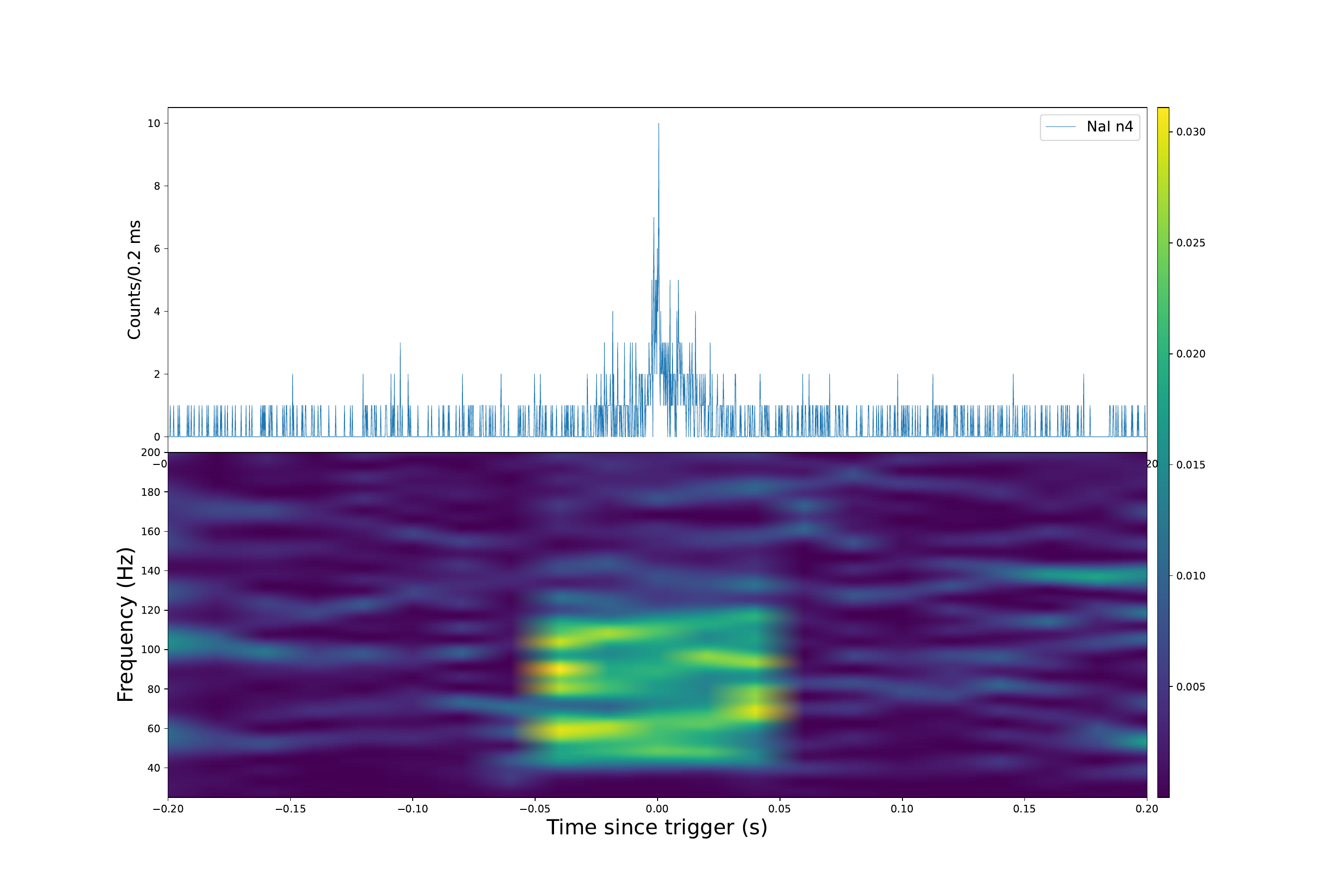}
	\includegraphics[width=0.48\linewidth]{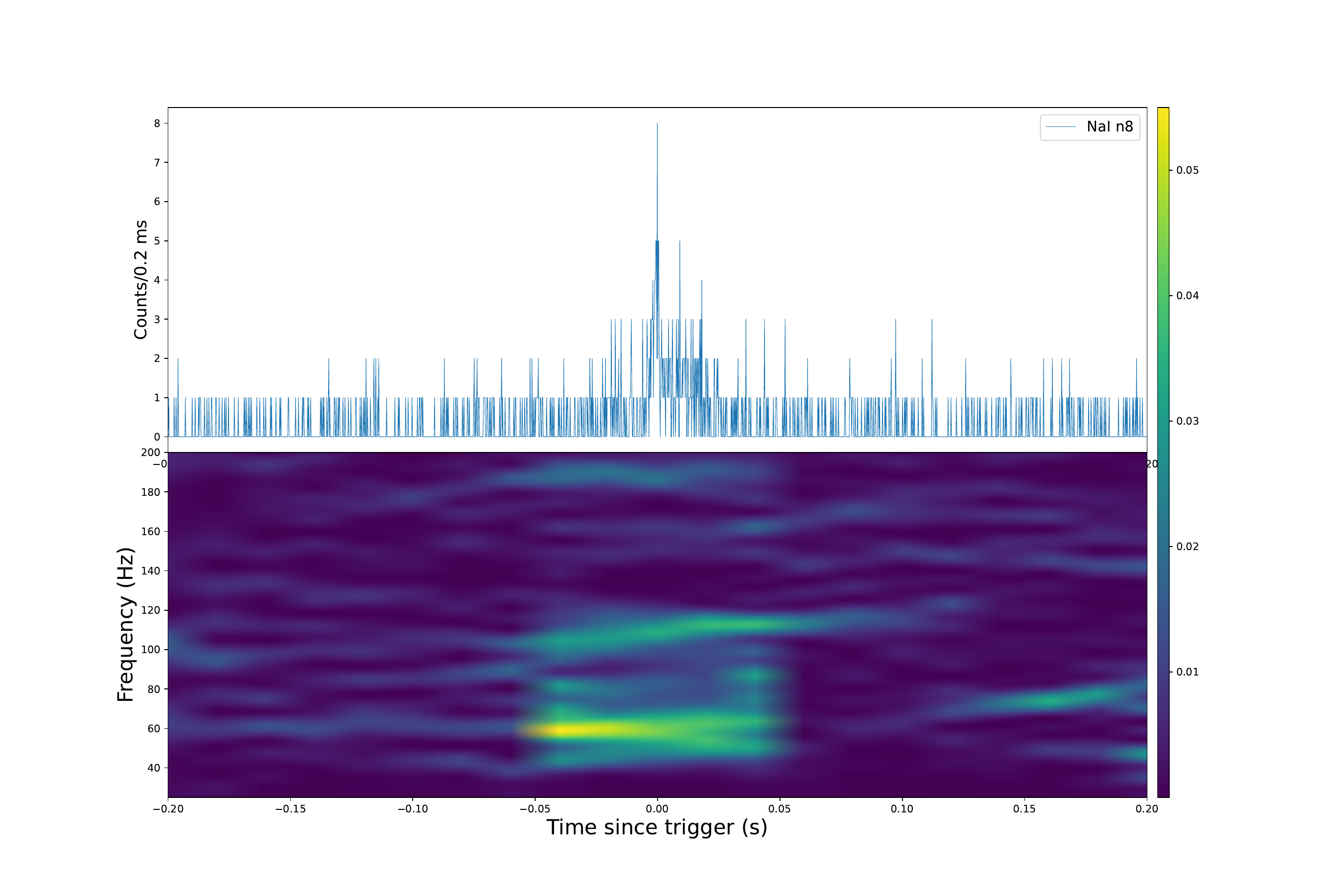}
	\caption{Lomb-Scargle periodogram analysis of SGR 150228213 between -0.2$-$0.2 s. Different panels denote analysis for detrended light curve from the combination of n4+n8 (top left), the combination of all NaI detectors (top right), detector n4 (bottom left) and detector n8 (bottom right). The energy range is 8-100 keV and the time resolution is 0.2 ms.}
	\label{LSP}
\end{figure}

To depict the variation of QPOs we discovered in burst light curve, we employed the Lomb-Scargle method \citep{1976Ap&SS..39..447L,1982ApJ...263..835S} to analyze the detrended light curve of SGR 150228213 in 8-100 keV.
Considering the weak signal-to-noise ratio in some untriggered detectors, we only analysed data from n4, n8, the combination of n4+n8 and the combination of all NaI detectors. 
A time window of length 0.1 s was used to produce Lomb-Scargle periodograms, which were combined into a spectrogram with time step of 0.2 ms.
The corresponding diagram is presented in Figure \ref{LSP}.
The analysis result for data of the combination of all NaI detectors is consistent with the QPOs detections, and the wide QPO at about 60 Hz appeared in the duration of about -0.06$-$0.05 s, the narrow QPO at about 110 Hz appeared in the duration of about -0.05$-$0.05 s.
Such QPOs are also visible in the results of n8 and the combination of n4+n8, and we can see that the most significant result exists on a single detector n8.
In addition, the result of n4 presented continuous power excess or nonsupport for the exsitence of QPOs.

From the results in Table \ref{pds_results2}, we can see that the significance is much lower in the shorter time interval centered on this burst, while we usually expect the opposite behavior if the QPOs were a real property of the burst.
However, as can be seen from the Figure \ref{LSP}, the most significant QPOs appeared at about -0.05 s, which may cause the lower siginicance in the shorter time interval.
In addition, the centroid frequencies of these two QPOs seem to have a relation of integral multiple, which indicates that the high frequency of 110 Hz (or 120 Hz) might be the second harmonic of the 55 Hz (or 60 Hz) fundamental.

\subsection {Gaussian Process analysis}
Gaussian processes (GPs) have been employed for searching QPOs in transient astrophysical events latest years \citep{2022ApJ...936...17H,2022ApJ...941..166X}, it models QPOs as a stochastic process on top of a deterministic shape, such deterministic shape can be understood as a mean model describe the overall trend of the burst light curve.
Since the QPOs at about 60 Hz and 110 Hz lies on the red-noise dominated part and was not confident enough based on the noise model, we can use GPs to verify whether such QPOs are generated from the red-noise process in time domain.

Following the procedure describe in \cite{2022ApJ...936...17H}, we defined the kernel function describing a QPO as
\begin{equation}
k_{\rm qpo}(\tau) = a\cos(2\pi f \tau)\exp(-c \tau),
\end{equation}
where $\tau$ is time constant, $a$ is the amplitude of the oscillation, $f$ is its frequency and $c$ is the inverse of the decay time of the QPO.
And the kernel function describing the red noise is defined as
\begin{equation}
k_{\rm rn}(\tau) = a\exp(-c \tau).
\end{equation}
As for the mean model function, since the unknown physical mechanism of SGR 150228213, we adopted three phenomenological which can describe the trend of light curves for gamma-ray bursts or flares, i.e., skewed Gaussians, skewed exponentials, and FRED models \citep{1996ApJ...459..393N,2015ApJ...810...66H,2022ApJ...936...17H}.
The significance of the QPO can be described by the the Bayes factor $BF_{\rm qpo}$, defined as
\begin{equation}
BF_{qpo} = \frac{Z(d|k_{\rm qpo+rn},\mu)}{Z(d|k_{\rm rn},\mu)},
\end{equation}
where $k_{\rm qpo+rn} = k_{\rm qpo}(\tau) + k_{\rm rn}(\tau)$ is the kernel function describes the QPO and the red-noise process with different $c$, $Z(d|k_{\rm qpo+rn},\mu)$ and $Z(d|k_{\rm rn},\mu)$ are the respective evidence in the QPO+red-noise and red-noise model, $\mu$ is the parameter of the mean function and $d$ is the data.

\begin{figure}
	\centering
	\includegraphics[width=0.32\linewidth]{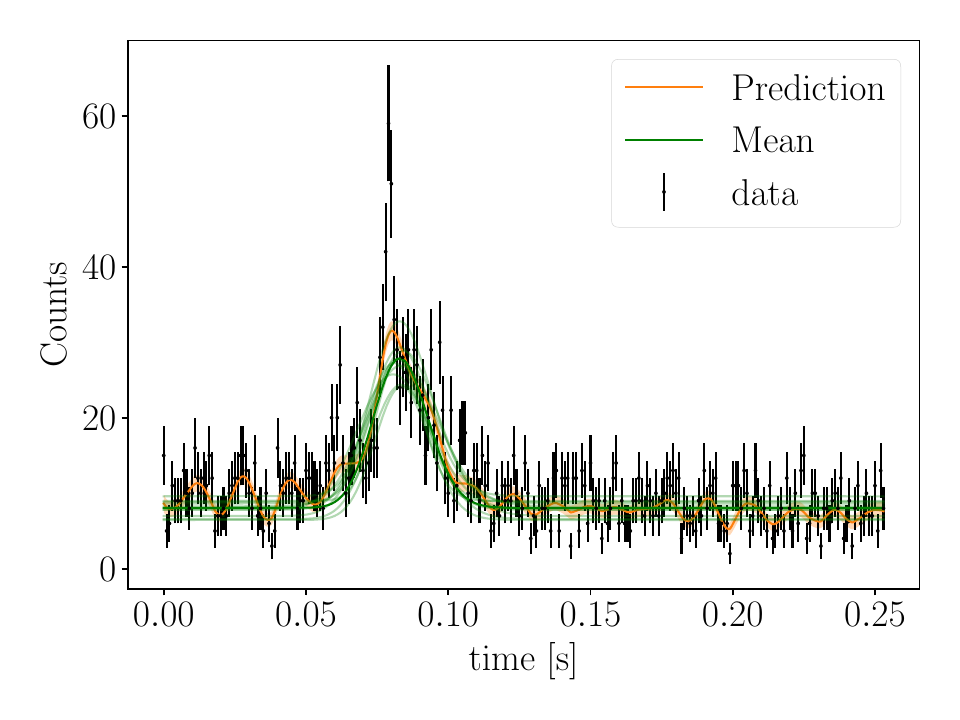} 
	\includegraphics[width=0.32\linewidth]{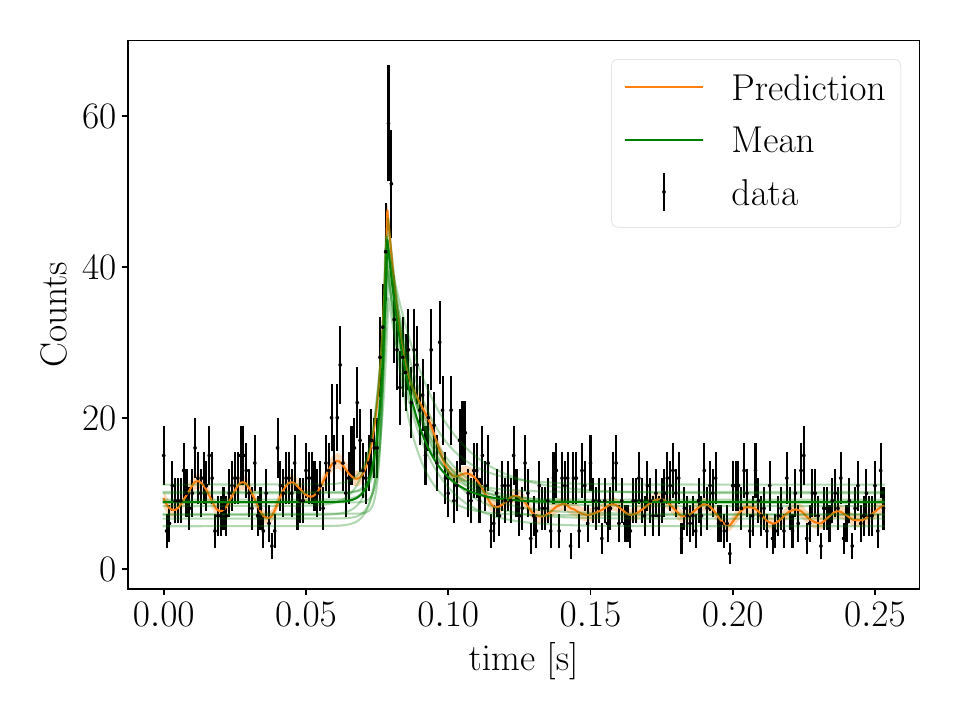}
	\includegraphics[width=0.32\linewidth]{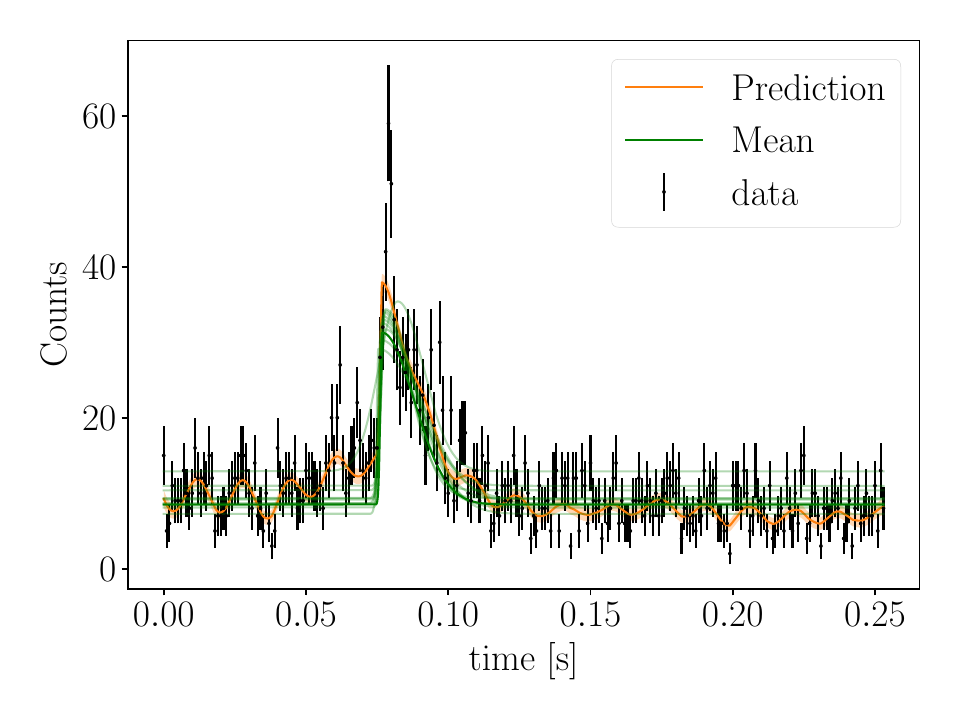}
	\includegraphics[width=0.32\linewidth]{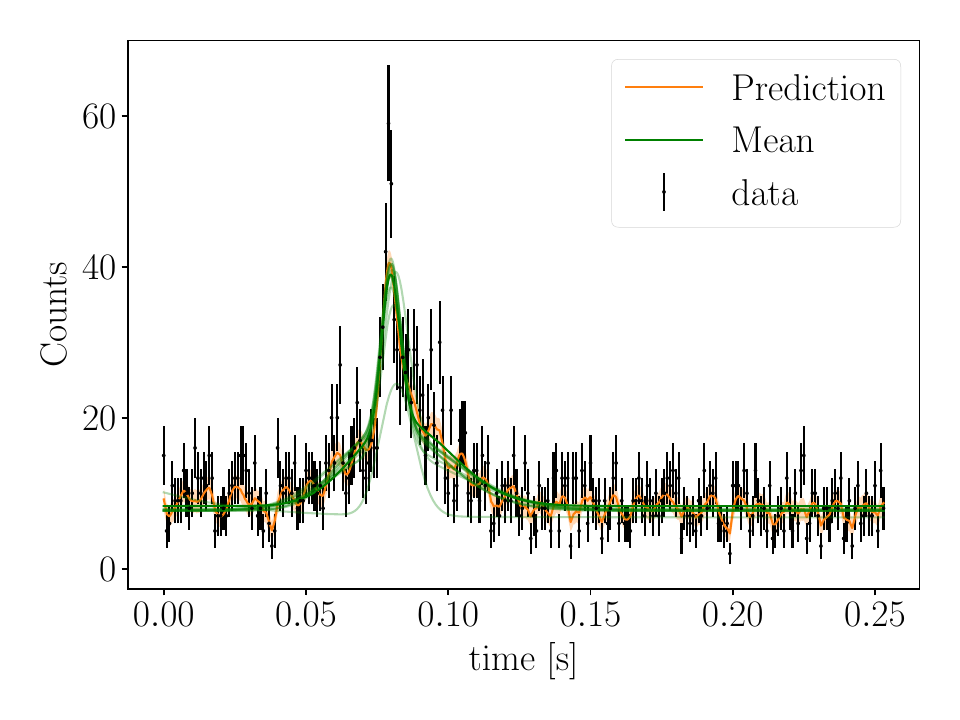}
	\includegraphics[width=0.32\linewidth]{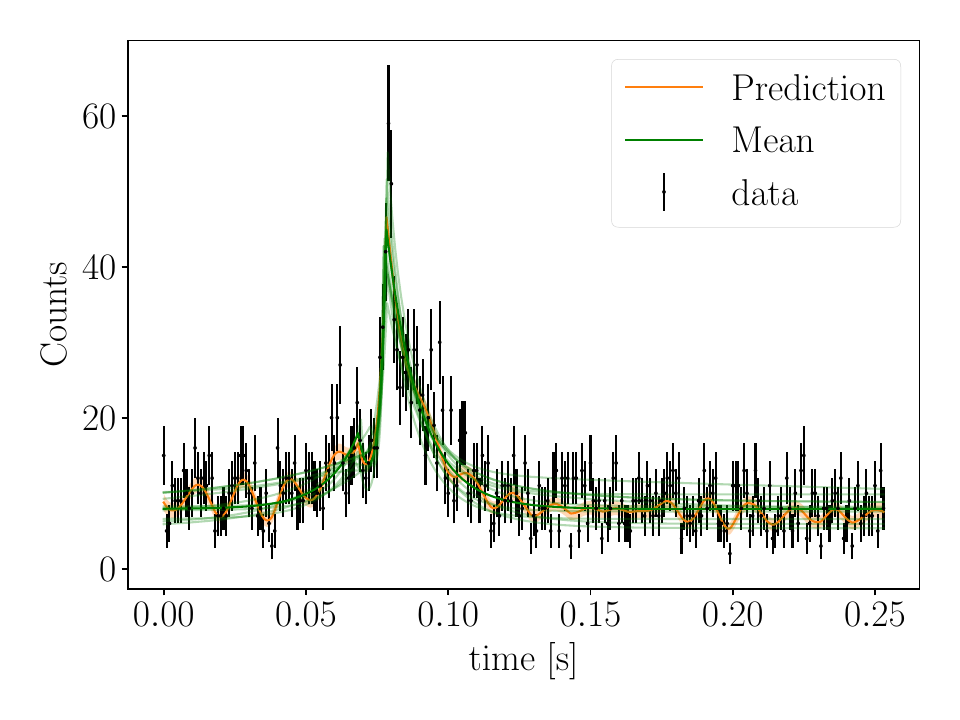}
	\includegraphics[width=0.32\linewidth]{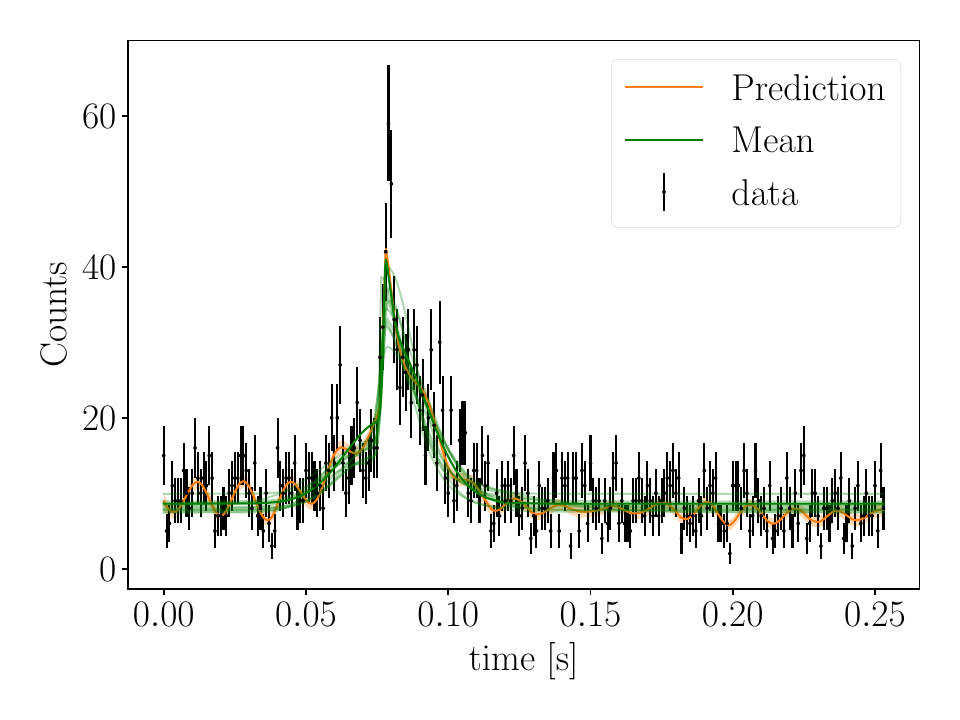}
	\caption{Gaussian process analysis of SGR 150228213 between -80$-$174 ms. Different panels denote analysis results using the $k_{\rm qpo+rn}$ kernel and different mean models, which contain one FRED (top left), one skewed exponentials (top middle), one skewed Gaussians (top right), two FRED (bottom left), two skewed exponentials (bottom middle) and two skewed Gaussians (bottom right). In each panel, black error bars denote the total light curve with 1 ms time resolution after zero correction, dark green line is the mean function from the maximum likelihood sample, light green lines denote 10 other samples from the posterior and orange line is the prediction based on the maximum likelihood sample and the 1$\sigma$ confidence band. The energy range is 8-100 keV and the time resolution is 1 ms.}
	\label{GPlc}
\end{figure}

\begin{figure}
	\centering
	\includegraphics[width=0.32\linewidth]{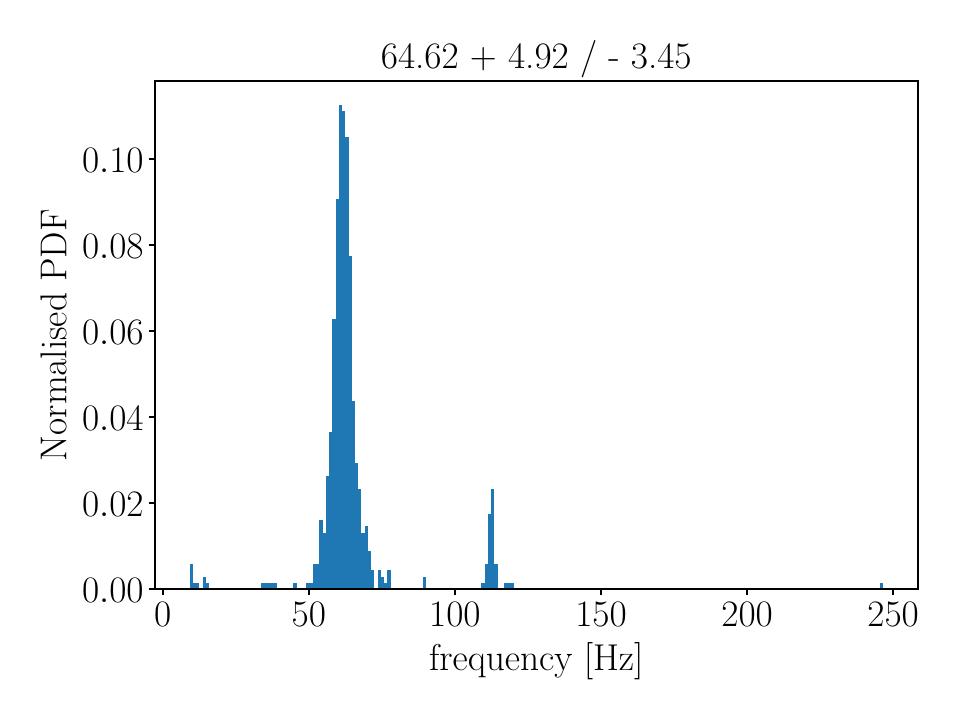} 
	\includegraphics[width=0.32\linewidth]{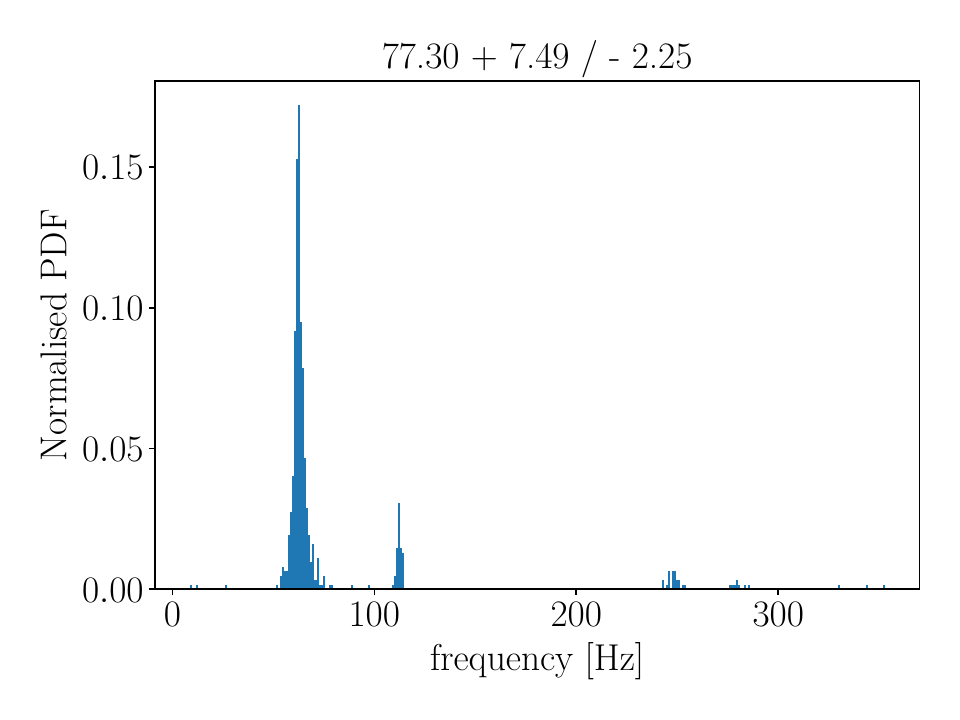}
	\includegraphics[width=0.32\linewidth]{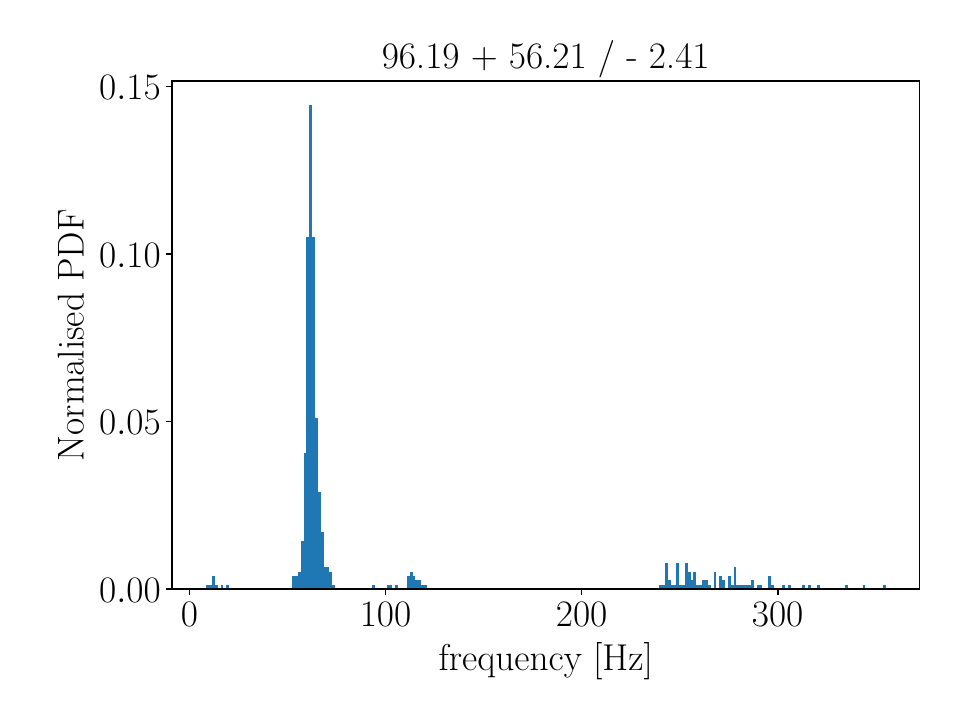}
	\includegraphics[width=0.32\linewidth]{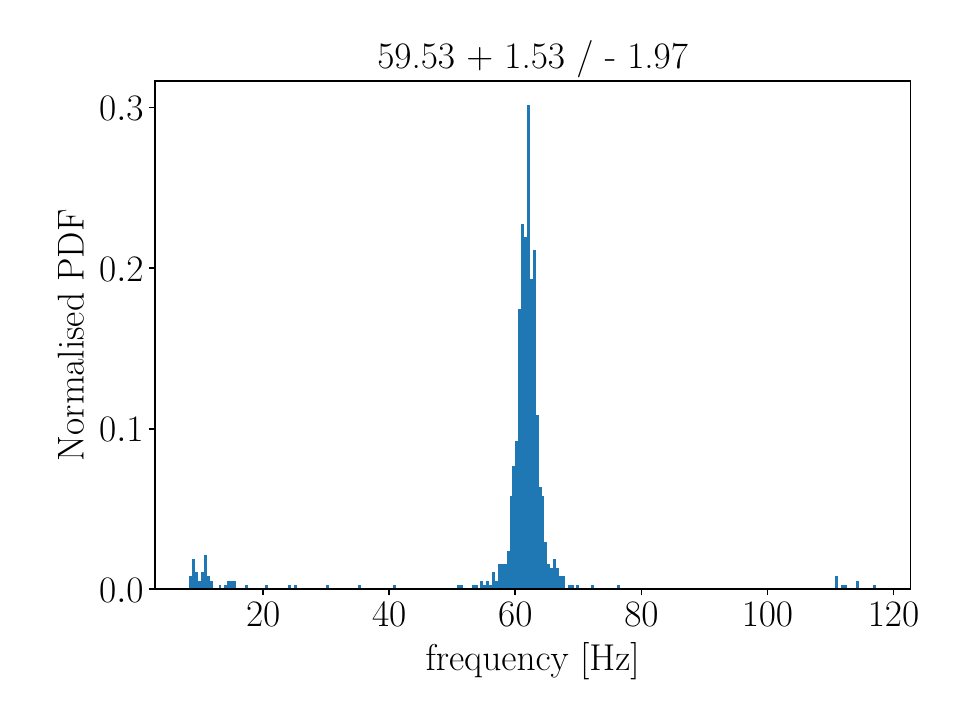}
	\includegraphics[width=0.32\linewidth]{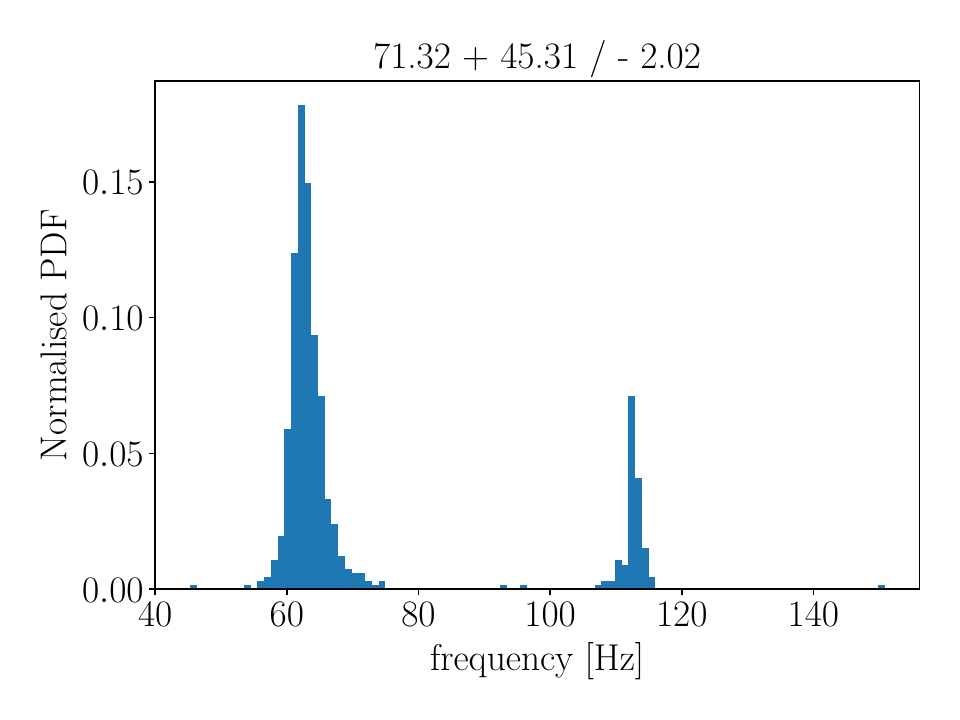}
	\includegraphics[width=0.32\linewidth]{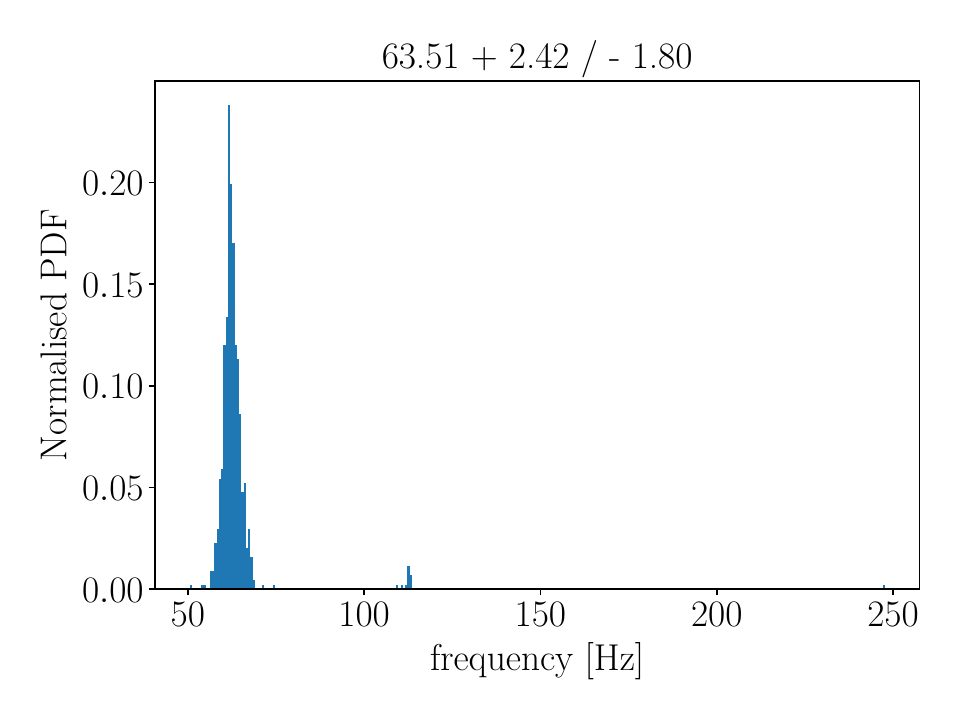}
	\caption{Frequency posterior distributions of SGR 150228213. Different panels denote analysis results under different mean models, which contain one FRED (top left), one skewed exponentials (top middle), one skewed Gaussians (top right), two FRED (bottom left), two skewed exponentials (bottom middle) and two skewed Gaussians (bottom right).}
	\label{GPp}
\end{figure}

In this section, we performed the GPs to the light curve data from the combination of all detectors in time interval of -80$-$174 ms to $T_0$ with 1 ms time resolution, and we made use of the publicly available code\footnote{https://github.com/MoritzThomasHuebner/QPOEstimation} of GPs released by \cite{2022ApJ...936...17H} to obtain the results.
According to the results, the QPO is disfavored under the mean models of one FRED ($\ln BF_{\rm qpo}$ = -1.8), two FRED ($\ln BF_{\rm qpo}$ = -0.97) and one skewed Gaussians ($\ln BF_{\rm qpo}$ = -0.78). 
And the QPO is favored under one skewed exponentials ($\ln BF_{\rm qpo}$ = 0.21), two skewed exponentials ($\ln BF_{\rm qpo}$ = 3.02), and two skewed Gaussians ($\ln BF_{\rm qpo}$ = 1.49).
The light curve under different mean models is presented in Figure \ref{GPlc}, and the frequency posterior is presented in Figure \ref{GPp}.
We found that the analysis results based on different mean models may (or not) have favorable to the existence of QPOs, and the skewed exponentials performed better than other models for burst profile if the Bayes factor is used for mean model comparisons.

In addition, the QPO frequency posterior for the SGR 150228213 is constrained in all models, and the results are consistent with the QPOs detection through frequency domain analysis.
As we concluded at Section \ref{sec:wavelet}, the QPO at about 110 Hz may be a second harmonic of the 55 Hz fundamental, and such conjecture seems also supported by the frequency distributions in Figure \ref{GPp}.
However, since the significance of such QPOs varies under different mean models, we reserved the results of frequency domain analysis as final judgment.
And we can see the potential of GPs for detecting QPOs in magnetar bursts, after all, the significance based on frequency domain analysis are usually recommended under the premise of infinitely long time series.

\section {Discussion on possible origins of SGR 150228213}\label{sec:discuss}
\subsection {SGR 150228213 as a magnetar burst from 4U 0142+61}
In the trigger report for SGR 150228213, $Fermi$/GBM attributed this burst to the activity of 4U 0142+61, for which the location of SGR 150228213 is close to this magnetar \citep{2015GCN.17508....1R}.
In addition, $Swift$ has detected a series of hard X-ray bursts from 4U 0142+61 $\sim$ 800 s before the trigger time of SGR 150228213 \citep{2015GCN.17507....1B}, these bursts have also been detected by $Fermi$/GBM.

4U 0142+61 is a prominent emitter in hard X-rays, optical and infrared \citep{2008A&A...489..245D,2004A&A...416.1037H}, it is the only magnetar with a debris disk but still debated whether it is an active gaseous one or a passive dust disk \citep{2006Natur.440..772W,2007ApJ...657..441E}.

For typical magnetar bursts, it is not clear of whether burst spectra are predominately thermal or non-thermal \citep{2011ApJ...739...87L,2012ApJ...749..122V}.
Table \ref{4u_spec} is the spectral fitting results for short burst from 4U 0142+61 detected by $Fermi$/GBM in 2015 collected from \cite{2017ApJ...835...68G}.
Here we select the 'preferred' model for each burst following the Bayesian information criterion (BIC) \citep{1978AnSta...6..461S}, the numerical value of which is calculated by
\begin{equation}
	{\rm BIC} = -2\ln \mathcal{L} + k\ln(d.o.f.),
\end{equation}
where $k$ is the number of parameters in the model, $d.o.f.$ is the data points used in fitting and $\mathcal{L}$ is the maximum likelihood.
When we compare the BIC of two models, if $\Delta$BIC $<$ 6, we consider there is no significant preference between both, if $\Delta$BIC $>$ 6, we prefer the model with smaller BIC \citep{1939thpr.book.....J,1998ApJ...508..314M}.

The 'preferred' model parameters for each burst are marked in bold in Table \ref{4u_spec}.
According to the energy spectrum fitting results, SGR 1502282123 is not significantly different from other bursts from 4U 0142+61. 
Moreover these short bursts from 4U 0142+61 detected in 2015 mostly have a harder energy spectrum than 'regular' short magnetars bursts (usually have $E_p$ below 50 keV in COMPT model fitting), which may indicate different physical origins to these bursts.
Unfortunately, we did not find any similar QPOs with SGR 150228213 in other bursts from 4U 0142+61, which may cause by the burst intensities are too low to provide sufficient significance for the potential QPOs.

Combined with the relationship to the active phase of 4U 0142+61 and location, 4U 0142+61 is undoubtedly the most likely origin of SGR 150228213.
If these QPO signal is not a false detection, this would be the first observation of QPOs in bursts from AXPs.
Considering the special feature of 4U 0142+61 itself, it may bring us new perspective for the burst mechanism of this magnetar.

\begin{table*}
	\setlength{\tabcolsep}{2.0pt}
	\caption{Spectral Parameters Comparison of SGR 150228213 with Bursts from 4U 0142+61$^{\textbf{a}}$}
	\label{4u_spec}
	\begin{center}
		\begin{tabular}{c c c c c c c c c}
			\hline\hline
			\multirow{2}{*}{Burst ID} & Start Time in UTC &  \multicolumn{3}{c}{BB+BB} & \multicolumn{3}{c}{COMPT} & Fluence \\
			\cmidrule(r){3-5}\cmidrule(r){6-8}
			& (2015 Feb 28) & $kT_1$ (keV)  & $kT_2$ (keV) & ${\chi}^2/d.o.f.$ & $\alpha$ & $E_p$ (keV) & ${\chi}^2/d.o.f.$ & ($10^{-8}$ erg cm$^{-2}$)\\
			\hline		
			1 & 04:53:25.023 & 7.9$\pm$1.9 & 19.0$\pm$4.6 & 83/64 & \textbf{-0.3$\pm$0.4} & \textbf{53.0$\pm$5.2} & \textbf{67/65} & 12$\pm$1\\
			
			2 & 04:53:35.195 & \textbf{2.8$\pm$0.8} & \textbf{17.6$\pm$1.2} & \textbf{55/64} & \textbf{-0.1$\pm$0.3} & \textbf{68.5$\pm$6.8} & \textbf{57/65} & 6$\pm$1\\
			
			3 & 04:57:21.307 & \textbf{5.1} & \textbf{21.5$\pm$2.3} & \textbf{51/65} & 0.4$\pm$0.7 & 82.0$\pm$12.0 & 66/65 & 9$\pm$1\\
			
			4$^{\textbf{b}}$ & 05:06:55.645 & 3.7$\pm$1.0 & 16.7$\pm$1.0 & 51/64 & \textbf{-0.2$\pm$0.3} & \textbf{60.6$\pm$4.6} & \textbf{47/64} & 29$\pm$2\\
			
			5 & 05:08:34.157 & 4.6$\pm$1.3 & 23.1$\pm$8.2 & 77/64 & {-1.9$\pm$0.8} & {29.7$\pm$101.0} & {54/65} & 3$\pm$1\\
			\hline
		\end{tabular}
	\end{center}
	\tablecomments{\textwidth}{\textbf{a.} Data collected from \cite{2017ApJ...835...68G}, these bursts are analyzed in 8-200 keV, only data from detectors with viewing angle $\leq$ 40$\degc$ to source is used. \textbf{b.} Corresponding to SGR 150228213.}
\end{table*}

\subsection {The relation between SGR 150228213 and FRB 180916}
Since SGR 150228213 is associated with FRB 180916 on location, we try to discuss the different origins of SGR 150228213 from a more interesting perspective.

FRB 180916 is an active repeating FRB source with a period of $\sim$ $16.35\pm0.15$ days and a 5 days phase window \citep{2020Natur.582..351C}, it was localized to a star-forming region in a nearby massive spiral galaxy at redshift z$\sim$$0.0337\pm0.0002$ \citep{2020Natur.577..190M}.
If this connection exists, SGR 150228213 may be a short GRB event generated from a newborn magnetar, which can also explain the highly active features of FRB 180916.

\subsubsection {Spectrum analysis and the $Amati$ relation}
If we treat SGR150228213 as a possible short GRB, we can use the $Amati$ relation \citep{2006MNRAS.372..233A} to check if it is correlated with the trend of short GRBs based on the energy spectrum analysis for it.
In this case, we used the COMPT model and the multi-color blackbody (mBB) model to fit the energy spectrum of SGR 150228213 in 8 keV$-$40 MeV, and check which model fits the burst better to compute the fluence of SGR 150228213.
We extract the source spectra, background spectra, and generate the instrumental response matrix from the detector n4, n8 and b0.
All spectra are fitted using {\tt Xspec} \citep{1996ASPC..101...17A}, we use the maximum likelihood for Poisson data with Gaussian background to estimate the best-fit parameters and choose the optimum model parameters through the MCMCs.

The COMPT model is defined as
\begin{equation}
	N(E) = KE^{\alpha} {\rm exp} [-(\alpha +2 )E/E_p ],
\end{equation}
where $K$ is the normalization factor, $\alpha$ is the photon index and $E_p$ is the peak energy in $\nu$$F_{\nu}$ spectrum.
The mBB model we used corresponds to the {\tt diskpbb} in {\tt Xspec}, and it is defined as \citep{2021ApJS..255...25I}
\begin{equation}
	N(E) = \frac{4 \pi E^2}{h^2 c^2} \left(\frac{K}{\zeta}\right) T_{p}^{(2/\zeta)} \int_{T_{min}}^{T_{p}} \frac{T^{\frac{-(2+\zeta)}{\zeta}}}{e^{(E/T)} - 1} dT,
\end{equation}
where $K$ is the normalization factor, $\zeta$ is power law index of the radial dependence of temperature ($ T(r) \propto r^{-\zeta}$), $T_{p}$ is the peak temperature in keV and $T_{min}$ is the minimum temperature of the underlying blackbodies and is considered to be well below the energy range of the observed data.

\begin{figure}
	\centering
	\includegraphics[angle=270,width=0.48\textwidth]{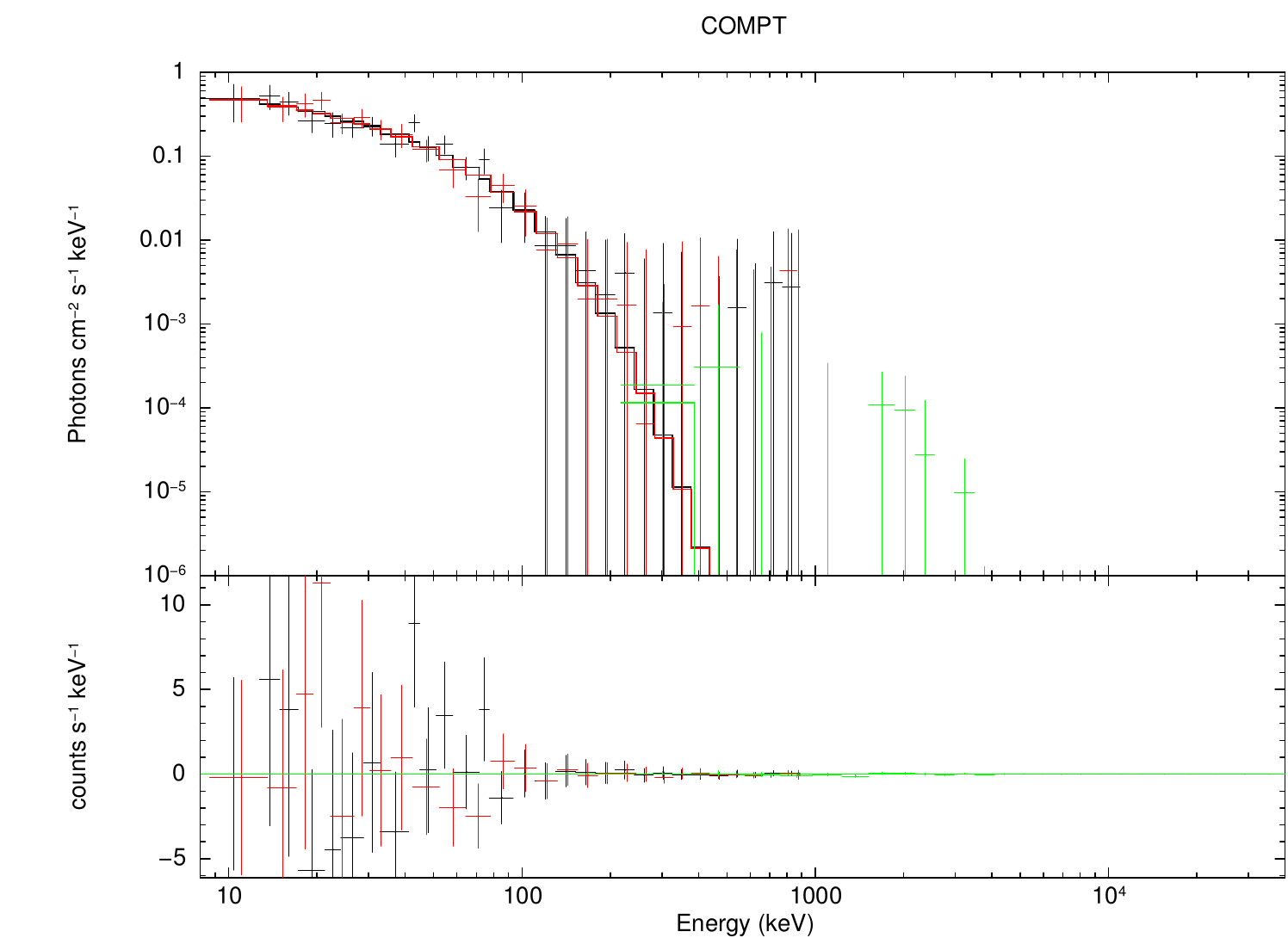}
	\includegraphics[angle=270,width=0.48\textwidth]{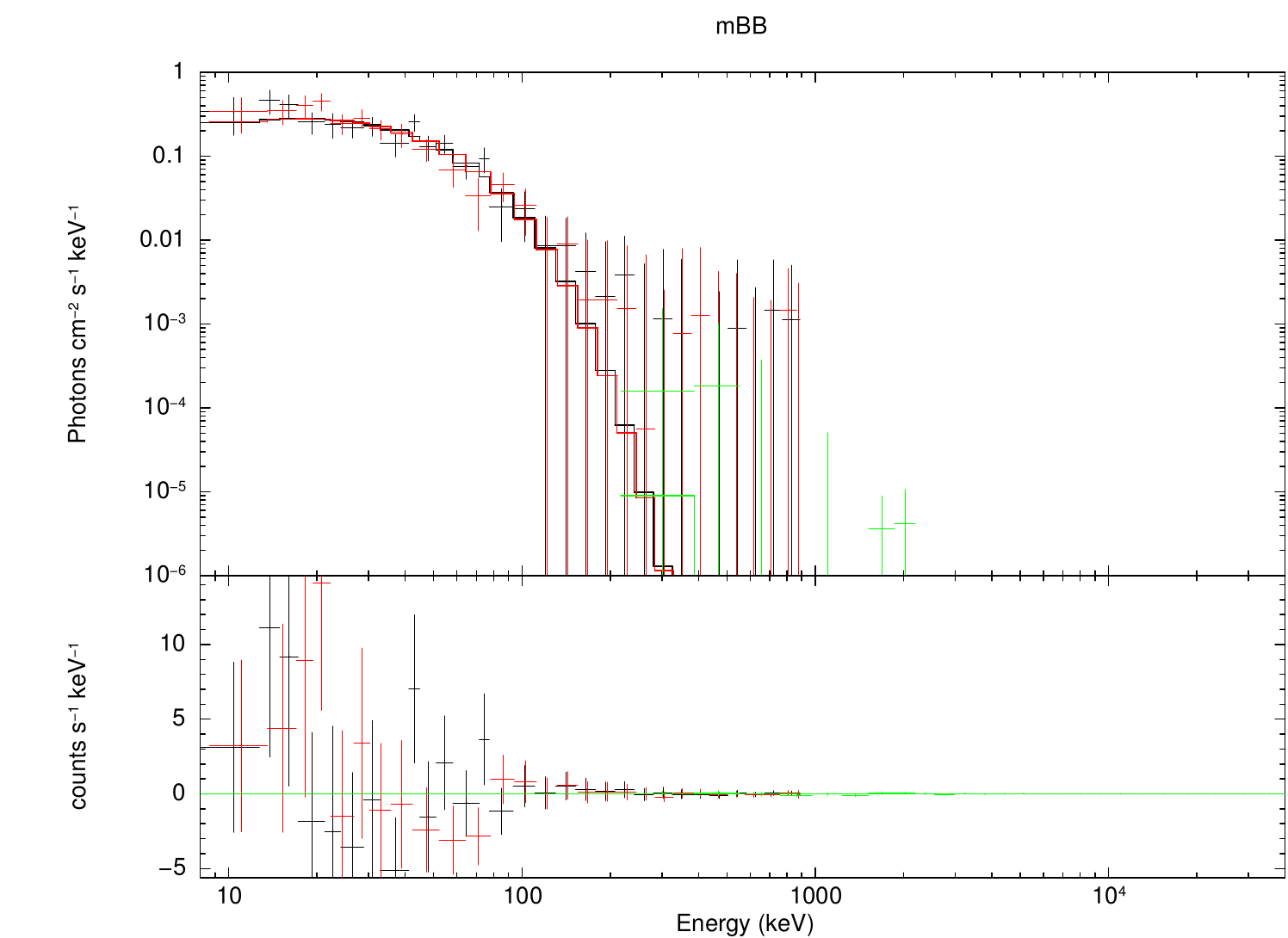}
	\includegraphics[angle=0,width=0.48\textwidth]{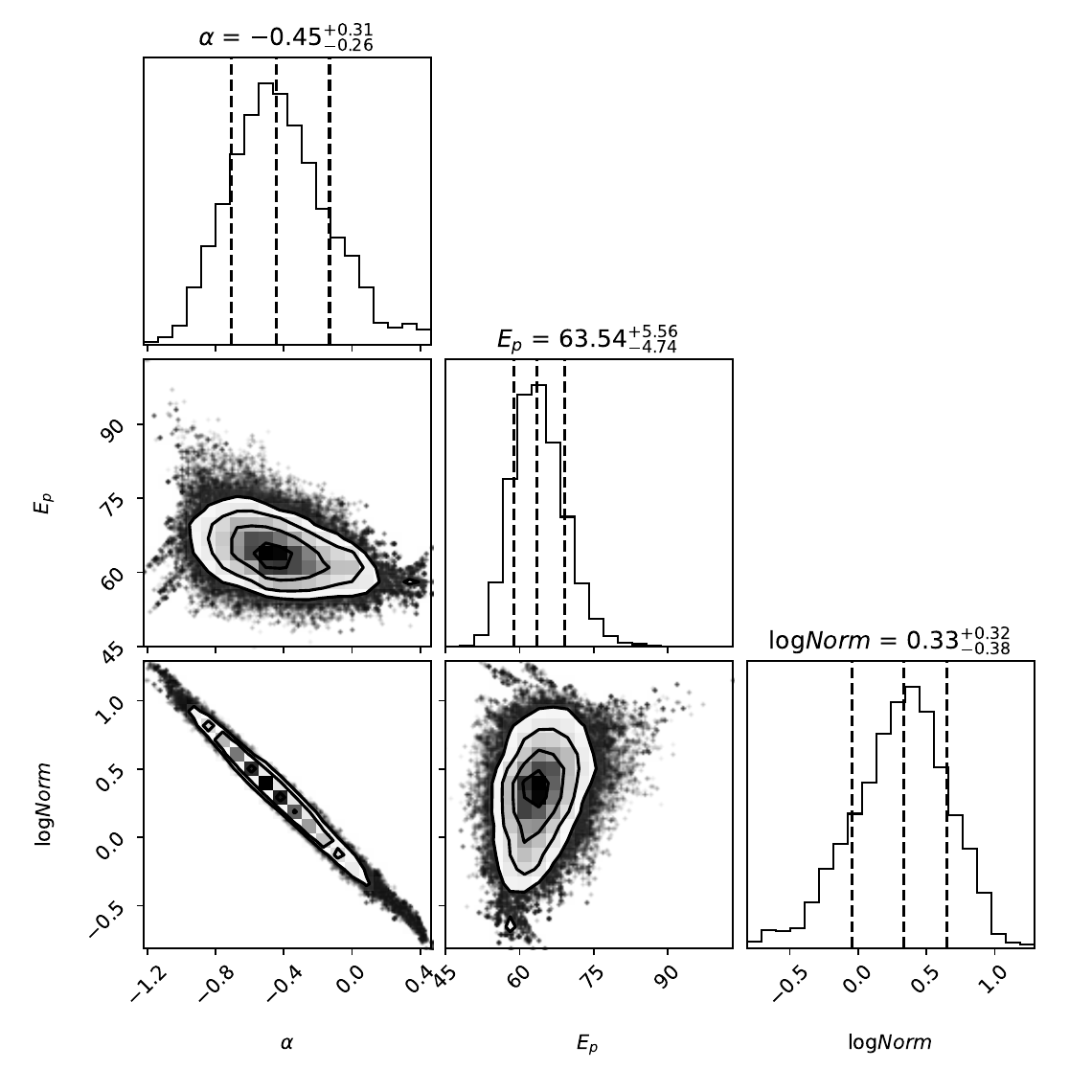}
	\includegraphics[angle=0,width=0.48\textwidth]{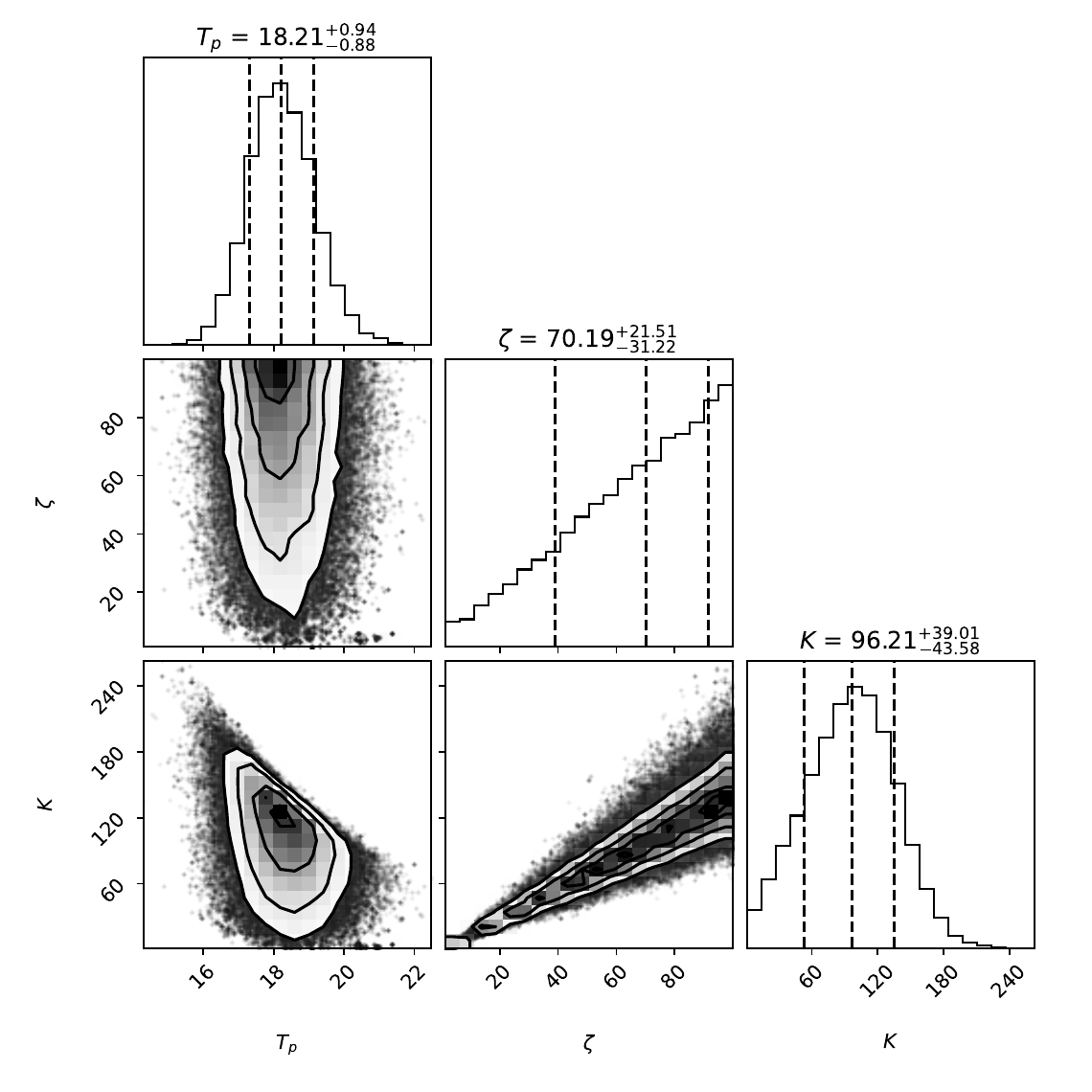}
	\caption{Time-integrated spectral fitting results in 8-40000 keV. Count spectrum in COMPT model (top left) and mBB model (top right) are drawn with their residuals. The likelihood map of free parameters in COMPT (bottom left) and mBB (bottom right) have been marked with their 1$\sigma$ uncertainties.}
	\label{spec}
\end{figure}

The spectrum of SGR 150228213 and model fitting results is presented in Figure \ref{spec}.
According to these results, the non-thermal origin of SGR 150228213 is still more supported and we can use the COMPT model fitting results to compute the $E_{\gamma,iso}$ of SGR 150228213 is $\sim$ $1.25\times10^{48}$ erg based on the redshift of FRB 180916.

According to the Amati relation, correlation between isotropic bolometric emission energy ($E_{\gamma,iso}$) and the rest-frame peak energy ($E_{p,z}$) could be written as
\begin{equation}
	\frac{E_{p,z}}{100\,keV} = C\left( \frac{E_{\gamma,iso}}{10^{52}\,erg} \right)^m,
\end{equation}
where C is around 0.8-1 and m is around 0.4-0.6.
This relation is initially found in long GRBs with known redshifts, but similar relations for short GRBs has also been found in later works \citep{2009ApJ...703.1696Z,2009A&A...496..585G}.

\begin{figure}
	\centering\includegraphics[width=\linewidth]{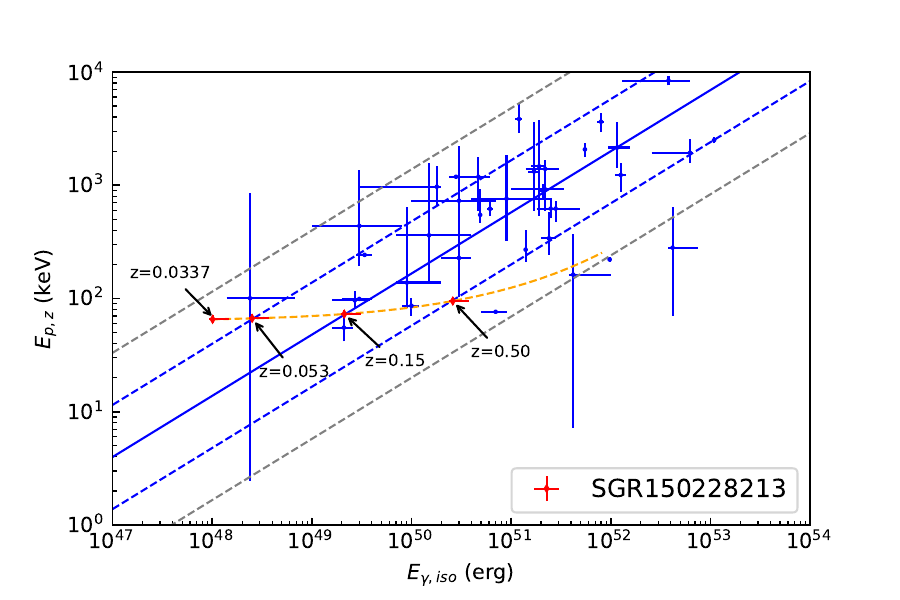} 
	\caption{SGR 150228213 in the $E_{p,z}$-$E_{\gamma,iso}$ correlation diagram of short GRBs. Blue solid line denotes the relation for short GRBs, blue and grey dashed lines denote the 1$\sigma$ and 2$\sigma$ regions. Orange dashed line denotes the SGR 150228213 position if it were taken redshifts from 0.0337 to 3. Red diamonds denote the SGR 150228213 position in z=0.0337 (the redshift of FRB 180916), z=0.15 (the 'best' position for SGR 150228213 in current correlation for short GRBs), z=0.053 and z=0.5. Other data of short GRBs are taken from \cite{2009ApJ...703.1696Z} and \cite{2018ApJ...859..160W}, the best correlation of short GRBs is taken as $\log E_p=(3.24\pm0.07)+(0.54\pm0.04){\rm log}(E_{\gamma,iso}/10^{52})$ \citep{2018NatCo...9..447Z}.}
	\label{amati}
\end{figure}

Figure \ref{amati} is the $E_{p, z}$-$E_{\gamma, iso}$ diagram of short GRBs, the position of SGR 150228213 in z=0.0337 is within the 1$\sigma$ and 2$\sigma$ error region of the distribution of short GRBs, and the 'best' redshift range for SGR 150228213 corresponding to short GRBs is z=$0.15^{+0.35}_{-0.097}$.

\subsubsection {Chance probability}
Apart from the possibility of verifying SGR 150228213 as a short GRB from the Amati relation, we need to estimate the chance probability of the  association between FRB 180916 and SGR 150228213.
However, the calculation may be suffer from some uncertainties.
Nevertheless, if we simply assumed that SGR 150228213 is a candidate for short GRB associated with FRB 180916. Following the methods in \cite{2020ApJ...894L..22W}, the chance probability of the association event may be calculated by 
\begin{equation}
P = 1 - \lambda^0\exp(-\lambda)/0! = 1 - \exp(-\lambda),
\end{equation}
where $\lambda = \rho S$ is the number of FRBs in the  region $S$ ($\approx [41252.96(1-\cos\delta R)]/2$).
The surface number density of our FRB samples is $\rho\approx 626/41252.96\approx0.015/\rm deg^2$. For the centering angular distance of FRB 180916 to SGR 150228213 $\delta R$ $\sim$ 0.4975$\degc$, one gets the chance probability $ \sim 1.16 \%$ \footnote{However,  if one takes $\delta R$ $\sim$ 5.45$\degc$ as the radius of the position error circle (with 90\% confidence) of SGR 150228213, then one get a much higher chance probability $\sim 24.69\%$ .}.
It can be seen that the chance probability of $\sim 1\%$ is relatively delicate, which implies the possibility of association, but it is not significant enough.
Therefore, combined with the physical analysis in the previous section, we leave open the possibility of a true association between SGR 150228213 and FRB 180916.

\section{Summary} \label{sec:sum}
After a Bayesian framework to the observed periodogram of SGR 150228213 based on the assumption that all broadband power in periodogram comes from the noise process without QPOs.
We detected a narrow QPO at 112.20 Hz with the width of 5.64 Hz and a wide QPO at 57.54 Hz with the width of 26.48 Hz in SGR 150228213, with a significance level of 0.0004 (corresponding to a confidence level $\simeq$ 3.35$\sigma$).

We have also discussed the possible origins of SGR 150228213, and consider it is most likely to come from the known magnetar 4U 0142+61.
If it indeed comes from 4U 0142+61, this would be the first detection of QPOs in bursts from AXPs, which may lead to new insights into the physical mechanisms of magnetar bursts.
However, we still do not rule out the possibility that it is a short GRB associated with FRB 180916.

\begin{acknowledgements}
We acknowledge the use of the $Fermi$ archive public data.
This work is supported by the National Natural Science Foundation of China (grant Nos. 12203013, 12273005 and U1938201),  China Manned Spaced Project (CMS-CSST-2021-B11) and  the Guangxi Science Foundation (Grant No. 2021AC19263).
\end{acknowledgements}

\bibliographystyle{raa}
\bibliography{paper}

\end{document}